\documentclass[reprint,aps,prx,twocolumn,superscriptaddress]{revtex4-2}
\usepackage{graphicx}
\usepackage{mathrsfs}
\usepackage{mathtools}
\usepackage{bm}
\usepackage{hyperref}
\usepackage{dcolumn}
\usepackage{amsmath}
\usepackage{amssymb}
\usepackage{color}
\usepackage{bbold}
\usepackage{braket}
\usepackage{xcolor}
\usepackage{changes}
\usepackage{ifpdf}

\newcommand{\bd}{{\mathbf d}}
\newcommand{\ba}{{\mathbf a}}
\newcommand{\br}{{\mathbf r}}
\newcommand{\bk}{{\mathbf k}}
\newcommand{\bq}{{\mathbf q}}

\newcommand{\bK}{{\mathbf K}}
\newcommand{\bR}{{\mathbf R}}
\newcommand{\bG}{{\mathbf G}}

\newcommand{\bt}{{\bm \tau}}

\begin{document}
\title{Mirror Symmetry Breaking and Lateral Stacking Shifts \\
in Twisted Trilayer Graphene}
\author{Chao Lei}
\thanks{C. L. and L. L. contributed equally to this work.} 
\affiliation{Department of Physics, The University of Texas at Austin, Austin, Texas 78712, USA}
\author{Lukas Linhart}
\thanks{C. L. and L. L. contributed equally to this work.} 
\affiliation{Department of Physics, The University of Texas at Austin, Austin, Texas 78712, USA}
\affiliation{Institute for Theoretical Physics, Vienna University of Technology, A-1040 Vienna, Austria}
\author{Wei Qin}
\affiliation{Department of Physics, The University of Texas at Austin, Austin, Texas 78712, USA}
\author{Florian Libisch}
\affiliation{Institute for Theoretical Physics, Vienna University of Technology, A-1040 Vienna, Austria}
\author{Allan H. MacDonald}
\affiliation{Department of Physics, The University of Texas at Austin, Austin, Texas 78712, USA}
\email{macd@physics.utexas.edu}

\begin{abstract}
    We construct a continuum model of twisted trilayer graphene using {\it ab initio} density-functional-theory calculations, and apply it to address twisted trilayer electronic structure.  Our model accounts for moir\'e variation in site energies, hopping between outside layers and within layers.
    We focus on the role of a mirror symmetry present in  
    ABA graphene trilayers with a middle layer twist.  The mirror symmetry is lost intentionally when a displacement field is applied between layers,
    and unintentionally when the top layer is shifted laterally relative to the bottom layer.  We use two band structure characteristics that are directly relevant to 
    transport measurements, the Drude weight and the weak-field Hall conductivity, and relate them via the Hall density 
    to assess the influence of the accidental lateral stacking shifts currently present in all experimental devices on electronic properties, and comment on the role of the possible importance of accidental lateral stacking shifts for superconductivity in twisted trilayers.
\end{abstract}

\maketitle

\section{Introduction}
Crystals with bandwidths that are small compared to electron-electron interaction scales feature strong electronic correlations, tending toward magnetism and, at integer total band fillings, Mott insulator states.  It has been found theoretically that the periodic moir\'e potential in twisted bilayer graphene (TBLG) slows low-energy electrons as the twist angle becomes small\cite{Lopes2007,Lopes2012}, with electron velocities vanishing at a discrete set of magic twist angles\cite{Bistritzer2011}. Interest in the flat moir\'e minibands of TBLG increased after the discovery of superconductivity and correlated insulating states near magic angle twists\cite{SC2018,Mott2018}. Recent experimental and theoretical work\cite{Codecido2019,Choi_2019,Jiang2019,Kerelsky2019,Xie2019,Chen2019,Lu2019,Yankowitz2019,Sharpe2019,Yankowitz2019,Sharpe2019,Xu2018,Koshino2018,Tarnopolsky2019,Seo2019,Lian2019,Song2019,Mora2019,Tomarken2019,Liu2019,Wolf2019,Chen_2020,Serlin_2020,Kerelsky_2019,Wong_2020,Kim_2020,Berdyugin_2020,Stepanov_2020,Uri_2020,Zondiner_2020,Utama_2020,Liu_2020,Lee_2019,Chittari_2019,Kang2018,Po2018,Guinea2019,Guinea_2018,Cea2020} has revealed a rich and detailed phenomenology that includes exotic topological and correlated insulating states\cite{Xu2018,Song2019,Ma2019,Mott2018}, stripe charge order\cite{Jiang2019} and ferromagnetism\cite{Sharpe2019}.

Twisted multilayer graphene systems have also attracted attention\cite{Morell2013,Liu2019,Shen_2020,Liu_2020,Ma2019,Mora2019,Carr_2020,Chebrolu2019,Cao_2020,Lado2020}, but are still relatively lightly explored.  In multilayers, flat bands tend to occur at larger twist angles\cite{Khalaf2019}, are less likely to yield energy gaps\cite{Mora2019}, and can support higher temperature superconductivity\cite{Park_2021,Haoeabg0399}. The low-energy electronic properties of multilayers depend on the orientation angle\cite{Mora2019} of each additional layer and on the relative stacking\cite{Khalaf2019}.  Furthermore, unlike the TBLG case, different layers have different chemical environments\cite{rickhaus2019_chemicalshift}, and will therefore differ in site energy.  Low-energy effective models are periodic ( with discrete exceptions ) only if the multilayer has only two distinct orientations among its layers. These complications severely challenge the effort to derive predictive, yet simple models for both structural and electronic properties.

\ifpdf
\begin{figure}[htp]
\includegraphics[width=0.9\linewidth]{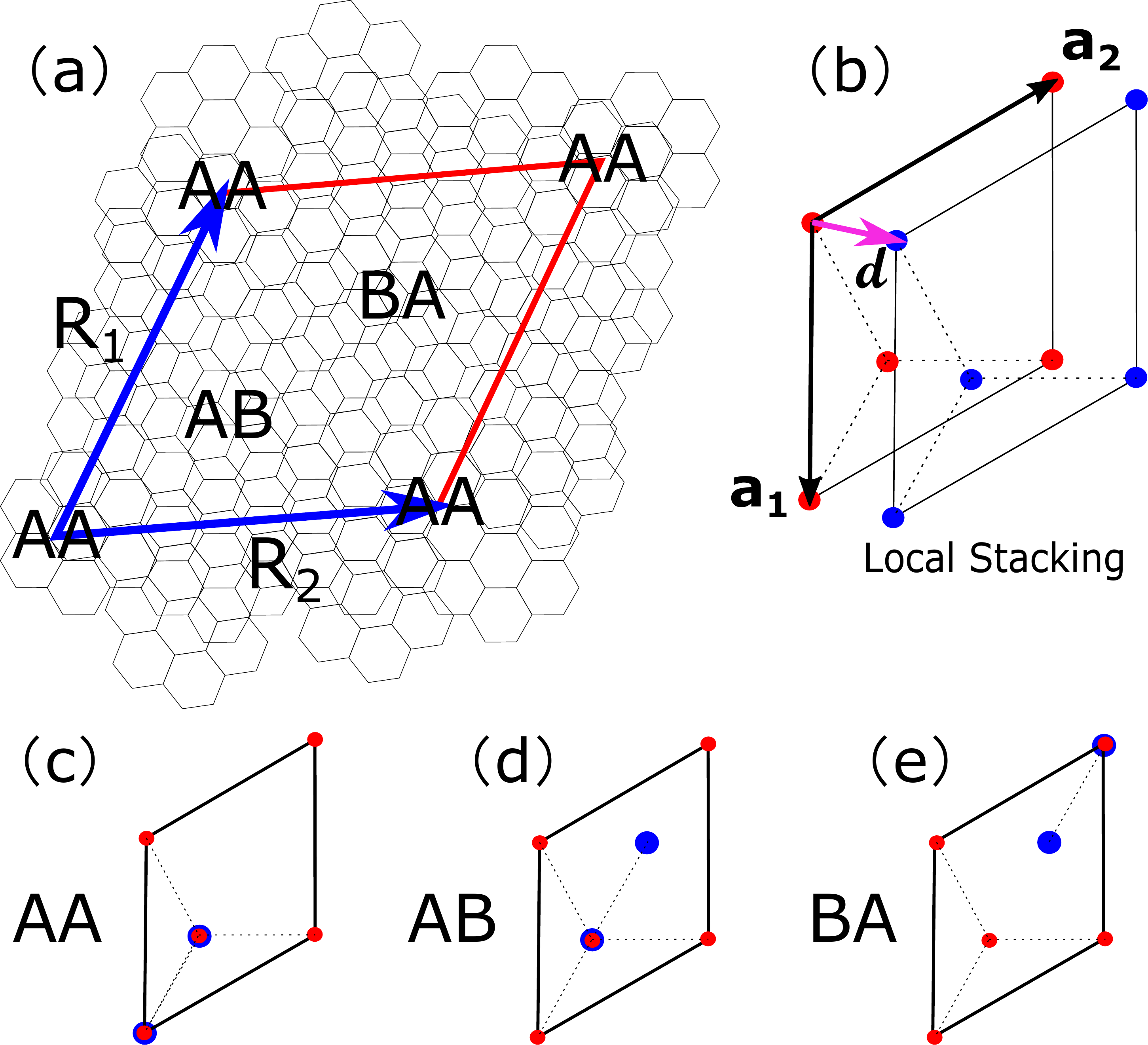}
\caption{Local stacking configurations in moir\'e lattices. (a) Primary moir\'e unit cell with lattice vectors $\bR_1$ and $\bR_2$. In the primary moir\'e unit cell local stacking between layers is dependent on the position within the primary moir\'e unit cell. (b) The local stacking value at a given position $\br$ within the moir\'e can be approximated by a local displacement vector $\bd$ within a pristine unit cell. The lattice vectors of the pristine graphene lattice are $\ba_1$ and $\ba_2$ as shown and the carbon atoms in the two layers are distinguished by red and blue circles, compare (c)-(e).}
    \label{fig:stacking}
\end{figure}
\fi

In this article we focus on twisted trilayer graphene (TTLG) in which two of the three layers are perfectly aligned.  At a qualitative level, these structures are interesting because they can yield cases in which flat bands with slow electrons and strongly dispersive bands that retain the isolated layer Dirac velocities are present simultaneously. We first derive low-energy continuum models from density-functional theory (DFT) calculations by modeling the local stacking (shown in Fig. \ref{fig:stacking}) for various high symmetry stacking and twist configurations.  These calculations therefore account not only for the modulation of closest layer hopping with moir\'e position, which is the dominant effect, but also for modulations of direct hopping between the outside layers, intralayer Dirac velocity, and site energies. Combining the effective continuum model and parameters from first-principle calculations we obtain a first-principle continuum model that not only provides a relative accurate modeling for twisted multilayer graphene, but also avoids large-scale first-principles calculations \cite{Lado2020,Cantele2020,Lucignano2019}. Indeed, the latter can only model compensated periodic moir\'e systems and cost tremendous computational resources,  especially when considering the effect of lateral stacking shifts and more layers. We find that the subdominant modulations play an essential role in determining the energetic alignments between the large and small velocity bands\cite{Taychatanapat_2011}.  More significantly we show that twisted trilayer graphene electronic structure has a qualitative sensitivity to relative lateral stacking shifts of the outer layers, which currently cannot be controlled experimentally, and to perpendicular electric fields. To provide physically relevant characterization of the influence of lateral stacking shifts on electronic structure, we calculate Drude weights, weak-field Hall conductivities, and Hall densities, which are experimentally accessible and can thus be used to provide a point of contact between theory and experiment.

In the absence of twists, ABA stacked trilayer graphene features an important mirror symmetry, which is preserved when only the middle layer is twisted.  The mirror symmetry decouples large velocity odd-parity and small velocity even-parity states\cite{li2019_trilayer,Khalaf2019}. A perpendicular electric field breaks the mirror symmetry while conserving C$_{2}T$ symmetry that allows the Dirac point degeneracies. More to the point, a lateral stacking shift of only the top layer of TTLG breaks both symmetries. Since it is -- for now --  not possible to stack layers with controlled lateral stacking shifts, realistic TTLG devices do not possess the mirror symmetry of the ideal middle-layer-twist structure. Indeed, lateral stacking shifts change the energetic alignment of the bands of the large- and small-velocity states. Combinations of perpendicular electric fields and lateral stacking shifts of the top layer change the electronic structure in a non-trivial way, which we characterize in terms of changes in the weak-field Hall conductivity, the Drude weight and the Hall density. 

The organization of this paper is as follow: In section~\ref{section_model} we introduce the continuum model used in this paper and explain how to obtain the model parameters from DFT calculations. In section~\ref{bands:stacking_dependence} we use the model to discuss and simplify the electronic band structures of several single twist TTLG stacking configurations, including the case of devices in which an outer layer is rotated and the other two layers are held in the AB stacking configuration. In section~\ref{band_characteristics} we introduce the experimentally relevant band characteristics, including the Drude weight, weak-field Hall conductivity and Hall density. In sections~\ref{section_electric_field} and \ref{section_stacking_shift} we discuss the dependence of the electronic structure, Drude weight, and weak-field Hall conductivity of middle-layer twist devices on electric field and lateral stacking shifts.  Here we find that lateral stacking shifts do have a strong influence on electronic properties, particularly electronic properties that are likely to be relevant for superconductivity, and that Hall density and Drude weight measurements can shed light on the lateral stacking shifts of particular devices.  Finally, in section~\ref{section_conclusionss} we summarize and discuss our results.

\section{Theoretical Methods}\label{section_model}

\subsection{Continuum model}

We describe the trilayer using a six-band continuum model that accounts for $\pi$-orbitals on both honeycomb sublattices of all three layers, allowing on-site energies and both intra-layer and interlayer coupling to vary spatially as the stacking changes on the moir\'e length scale. The approach we take in this paper can be generalized from the trilayers we consider to arbitrary graphene multilayers, all of which have Hamiltonians $\mathcal{H}$ that can be partitioned as follows:
\begin{equation}\label{equ:phenomen}
    \mathcal{H} = 
    \int d\br \big[ \mathcal{H}_{D} +
    \mathcal{H}_{s}(\br) +
    \mathcal{H}_{t}(\br) +
    \mathcal{T}(\br) \big] ,
\end{equation}
where $ \mathcal{H}_{D}$ is a valley-projected isolated layer Dirac Hamiltonian for layer $l$ with orientation:
\begin{equation}\label{eq:Dirac}
    \mathcal{H}_{D} = \sum_{l\alpha\beta}
    h^{\alpha\beta}(-i\partial_{\br},\theta_l)
    c_{l\alpha}^{\dagger} c_{l\beta} .
\end{equation}
$ h(-i\partial_{\br},\theta_l) = h(\bk,\theta_l)|_{\bk \rightarrow -i\partial_{\br}} $ is given explicitly in App.~\ref{app:pristine}. In Eq.~\ref{eq:Dirac} $\alpha,\beta$ are sublattice labels. $\mathcal{H}_{s}(\br) $ accounts for corrections to on-site energies due to the moir\'e pattern,
\begin{equation}\label{equ:onsite}
    \mathcal{H}_{s}(\br) = \sum_{l,\alpha} 
    \epsilon^{l}_{\alpha}(\mathbf{r})
    c^{\dagger}_{l\alpha}(\br)
    c_{l\alpha}(\br), 
\end{equation}
and $ \mathcal{H}_{t}(\br) $ for corrections to the coupling within and between aligned layers
\begin{equation}
    \mathcal{H}_{t}(\br) = 
    \sum_{ll^{\prime},\alpha \beta^{\prime}}
    t^{ll^{\prime}}_{\alpha\beta^{\prime}}
    (\mathbf{r})
    c^{\dagger}_{l\alpha}(\br)
    c_{l^{\prime},\beta^{\prime}}(\br),
\end{equation}
{\it i.e.}, to spatial variations of the Fermi velocity and coupling variations between aligned layers. Finally, $\mathcal{T}(\br)$ is the term that captures tunneling between layers that have different orientations, the term which most prominently alters single-layer electronic structure in moir\'e multilayers. The interlayer tunneling Hamiltonian has the same form as that of a tight-binding model in which the coupling between sites is a function of lateral stacking shifts\cite{Bistritzer2011}:
\begin{equation}\label{equ:tunneling_amp}
    \mathcal{T}(\br) = \sum_{l \ne l^{\prime}
    \alpha\beta^{\prime};n}
    \omega_{\alpha\beta'}^{ll',\bG_n}
    e^{-i\bq_n^{ll^{\prime}} \cdot \br}
    c_{l\alpha}^{\dagger}(\br)
    c_{l^{\prime}\beta^{\prime}}(\br),
\end{equation}
where $l(l^{\prime})$ labels layer and $\bq_n^{ll^{\prime}} $ is the Dirac point momentum difference between the $l$ and $l^{\prime th}$ layer, with
\begin{equation}\label{equ:cond}
    \bq_n^{ll^{\prime}}  = 
    \bK_l - \bK_{l^\prime} + 
    \tilde{\bG}_n^{ll^{\prime}}.
\end{equation}

Here $\tilde{\bG}_n^{ll^{\prime}} \approx - \theta \hat{z} \times \bG_n^{ll^{\prime}}$ is the corresponding difference between the reciprocal vectors that connect equivalent downfolded Brillouin-zone corners $\bK_{l}(\bK_{l^{\prime}})$ \cite{Bistritzer2011}.  Since the coupling elements that have equal magnitude momentum boosts $|\bq_{n}^{ll^{\prime}}|$ are related to each other by symmetry, we group the $\bq_{n}^{ll^{\prime}}$ into "shells" of equal length.  In this approach it is natural to explicitly exhibit the dependence on sublattice due to the difference in corresponding $\pi$-orbital Wannier functions positions by letting $ \omega_{\alpha\beta^{\prime}}^{ll^{\prime},\bG_n} \to w_{\alpha\beta^{\prime}}^{ll^{\prime}} e^{i\bG_n^{ll^{\prime}} \cdot (\bt_{\alpha} - \bt_{\beta^{\prime}} )} $, where $ \bt_{\alpha} $ specifies the $\alpha$ sublattice position in the $l^{th}$ layer and $ \bt_{\beta^{\prime}} = \bt_{\beta} + \bd $ specifies a sublattice position in the $l^{\prime th}$ layer in which $\bd$ is shown in Fig. \ref{fig:stacking} (b). 
 
In TBLG this approach can be successfully applied on a purely phenomenological basis, because the number of required parameters is small.  In TTLG, however, the difference in chemical environment between the inside and outside layers, and the possibility of tunneling directly between top and bottom layers causes a proliferation in parameters and guidance from \textit{ab initio} theory is needed.

\subsection{Ab initio calculations}

\ifpdf
\begin{figure*}
    \includegraphics[width=0.95\textwidth]{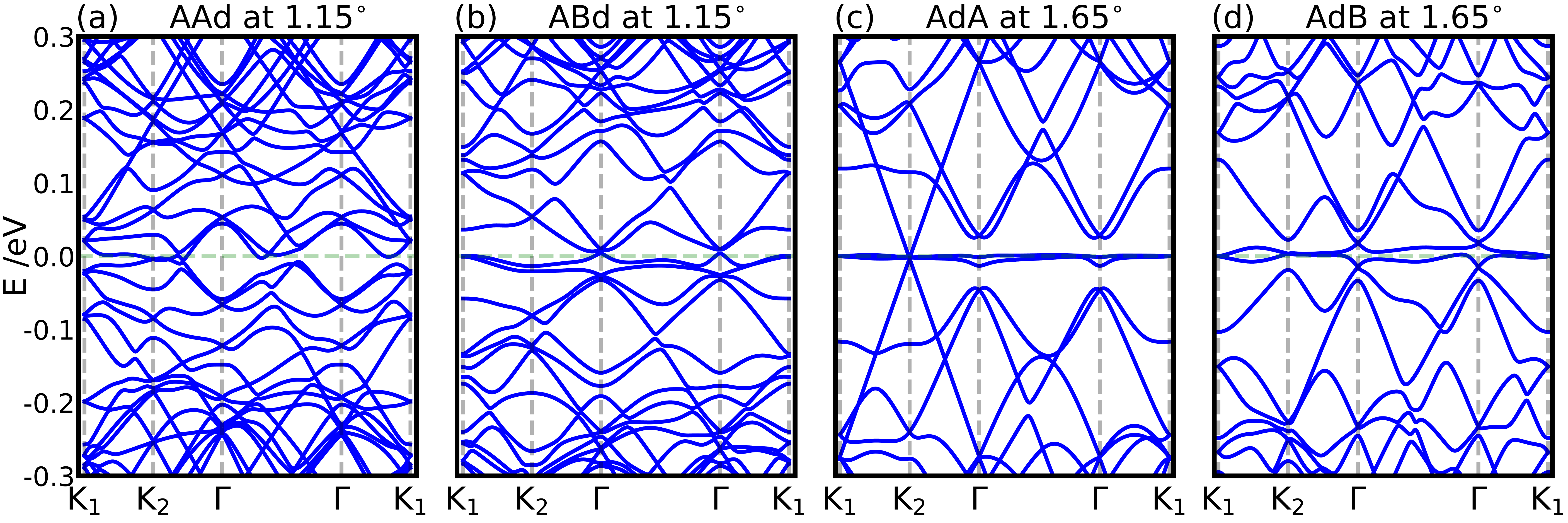}
    \caption{Continuum model moir\'e bandstructures for four different TTLG at the largest magic angle including all {\it ab initio} derived coupling elements. (a) top-layer twisted trilayer with the bottom two layers in AA stacking (AAd); (b) top-layer twisted trilayer with the bottom two layers in AB stacking (ABd); (c) middle-layer twisted trilayer with the outer two layers in AA stacking (AdA).  This structure has an emergent mirror symmetry; (d) middle-layer twisted trilayer with the outer two layers in AB stacking (AdB).}
        \label{fig:abinit_bs}
\end{figure*}
\fi

Closely following an approach introduced by Jung~et~al.~\cite{Jeil2014}, we employ the approximation that the small twist angle moir\'e Hamiltonian mainly depends on the local coordination between the layers. We are interested in the case in which one subset of layers is twisted relative to the others by an angle $\theta$. If the twist is rigid, the twisted layers are displaced relative to the untwisted layers by (compare Fig.~\ref{fig:stacking})
\begin{equation}
  \bd(\br) \approx
  \theta \, \hat{z} \times \br.
\end{equation}
Truncating the Hamiltonian to $\pi$-orbitals, the local Hamiltonian $\mathcal{H}(\br)$ can be expressed in terms of the displacement vector $\bd(\br)$,
\begin{equation}
  H(\bk,\bd(\br)) = \sum_{ll',\alpha\beta^{\prime}} \langle l\alpha\bk | \mathcal{H}(\br) | l^{\prime}\beta^{\prime}\bk  \rangle \; | l\alpha\bk \rangle  \langle l^{\prime}\beta^{\prime}\bk|.
\end{equation}
Here $l$ labels layer, $\alpha(\beta^{\prime})$ labels sublattice sites within each layer's honeycomb, and $\bd$ is the displacement of the twisted layers. The Hamiltonian as defined above is periodic in $\bd$ when translated by pristine lattice vectors. By Fourier transforming the Hamiltonian matrix elements subject to such a translation condition via,
\begin{eqnarray}\label{e:Hllprime}
    H^{ll^{\prime}}_{\alpha\beta}(\bk,\bd(\br)) 
    &=& \sum_{\bG}
    \hat{H}^{ll^{\prime}}_{\alpha\beta}(\bk,\bG)
    e^{i \tilde{\bG} \cdot \mathbf{r} },  
\end{eqnarray}
and using the identity $\tilde{\bG} \cdot \br =  \bG_n \cdot (-\theta \hat{\bm z} \times \br) = -\bG \cdot \bd(\br)$ one can approximate the $\br$ dependence of the Hamiltonian with just a few Fourier components. This approach will only be successful if $|H^{ll^{\prime}}_{\alpha\beta}(\bK,\bG)|$ drops to zero sufficiently quickly with increasing $\bG$.  

We find that this condition is fulfilled due to the smoothness of the parameter variation obtained from the sampled configuration space. Indeed, using Eq.~\ref{e:Hllprime} it is possible to identify the expansion coefficients $\hat{\epsilon}^{l}_{\alpha,\bG}$, $\hat{t}^{ll^{\prime}}_{\alpha,\beta,\bG}$ and $\omega_{\alpha\beta}^{ll^{\prime},\bG}$ of Eqs.~\ref{equ:onsite}-\ref{equ:tunneling_amp} with the corresponding $\hat{H}^{ll^{\prime}}_{\alpha,\beta}(\bK,\bG)$ \cite{Jeil2014}. We thus obtain a moir\'e band model from {\it ab initio} calculations performed on pristine cell graphene multilayer structures that has the same structure as the continuum model outlined in section~A. For further details on the \textit{ab initio} calculations we refer to App.~\ref{ab_intio_details}.\\

\begin{table}
    \begin{tabular}{c||c|c|c|c|c|c}
        \hline
        Shell& 
        $\omega^{b,m}_{A,A}$ & 
        $\omega^{b,m}_{A,B}$ &
        $\omega^{m,t}_{A,A}$ & 
        $\omega^{m,t}_{A,B}$ &
        $\omega^{b,t}_{A,A}$ & 
        $\omega^{b,t}_{A,B}$ \\ \hline \hline
        &\multicolumn{6}{c}{AAd} \\ \hline
        1$^{st}$ & - & - & 91 & 106   & 2 & 3 \\
        2$^{nd}$ & - & - & -9 & 6     & 0 & 0 \\
        3$^{rd}$ & - & - & -5 & 4(47) & 0 & 0 \\
        \hline
        &\multicolumn{6}{c}{ABd} \\ \hline
        1$^{st}$ & - & - & 90 & 105   & 3 & 3 \\
        2$^{nd}$ & - & - & -9 & 5     & 0 & 0 \\
        3$^{rd}$ & - & - & -4 & 5(56) & 0 & 0 \\
        \hline
        &\multicolumn{6}{c}{AdA} \\ \hline
        1$^{st}$ & 93 & 105  & 93 & 105   & - & - \\
        2$^{nd}$ & -8 &  4   & -8 & 4     & - & - \\
        3$^{rd}$ & -5 & 4(50)& -5 & 4(50) & - & - \\
        \hline
        &\multicolumn{6}{c}{AdB} \\ \hline
        1$^{st}$ & 94 & 107   & 91 & 105   & - & -\\
        2$^{nd}$ & -8 & 6     & -9 & 5     & - & -\\
        3$^{rd}$ & -4 & 5(49) & -4 & 5(43) & - & -\\
        \hline
    \end{tabular}
    \caption{Interlayer tunneling parameters $\omega^{ll^{\prime},\mathbf{G}}_{\alpha\beta}$ for different single-layer-twist trilayer structures.  The four stacking configurations (AAd, ABd,AdA, and AdB) are specified by listing the layers from bottom to top, with the twisted layer labelled $d$ to suggest the displacements used in the DFT calculations to represent twist locally. Each shell corresponds to momentum boost $\bq_{n}^{ll^{\prime}}$ shells containing members of equal length (see Eq.~\ref{equ:cond}). More explicitly to $\bG_n^{ll^{\prime}} = i\cdot\mathbf{b}_{1} + j\cdot \mathbf{b}_{2} $ with $(i,j)$ $\in$ $\{(0,0),$ $(1,-1),$ $(1,0)\}$ for the 1$^{st}$ shell, $(i,j)$ $\in$ $\{(0,1),$ $(0,-1),$ $(2,-1)\}$ for the 2$^{nd}$ shell, and $(i,j)$ $\in$ $\{(2,-2),$ $(1,-2),$ $(-1,0)\}$ for the 3$^{rd}$ shell. Non-zero phases are denoted in brackets. The band parameters are given in meV units and b,m,t label the bottom, middle and top layer respectively.  The smaller band parameters that account for site-energy variation and hopping between aligned layers are listed in the appendix.
    }
    \label{tbl:coupling_elements}
\end{table}

A detailed analysis of the resulting band parameters reveals that many arguments advanced for the TBLG case also hold for the trilayer case. In line with TBLG and previous publications on TTLG\cite{Khalaf2019,Mora2019} the dominant contribution to the interlayer coupling strength stems from the first shell interlayer coupling elements which are of similar magnitude as the values for TBLG. The first shell here denotes the $\bq_n$ that connect the two layers at the first Brillouin zone edges at the vicinity of the $\bK$ point (see Table~\ref{tbl:coupling_elements}). However, the second shell contributions are not completely negligible in the TTLG case (see~Table~\ref{tbl:coupling_elements}). Equally subtle and relevant is the energy alignment between the layers, and the spacial variation of on-site energies and Fermi velocities on the length scale of the moir\'e, and the tunneling elements that coupling the outer two layers (see~Tables~\ref{tbl:intralayer_coupling_elements}~and~\ref{tbl:aligned_coupling_elements}).

\section{Twisted Trilayer Electronic Structure}\label{section_bands}

Armed with realistic single-particle trilayer Hamiltonians, we are now in a position to address electronic structure. The most important difference compared to the simple TBLG case is that electronic properties are in most cases sensitive to translations of individual aligned layers \cite{Khalaf2019,li2019_trilayer}. We focus on four special cases: a middle-layer twisted trilayer with the outer layers in either AA stacking (AdA), or AB stacking configuration (AdB), and top-layer twisted trilayers with the other two layers in either AA (AAd) or AB stacking (ABd) (see~Fig.~\ref{fig:stacking}). Our results show that the moir\'e band-structure of twisted trilayers not only depends on parameters that are in principle experimentally controllable such as the twist-angle or a gate-controlled displacement field, but also on one variable beyond current experimental control, namely single-layer lateral displacements on the atomic scale.  

\subsection{Trilayer Stacking Dependence}\label{bands:stacking_dependence}

We first contrast the continuum model bandstructures for the four different single-twist trilayers (compare Fig.~\ref{fig:abinit_bs}), each calculated at a twist angle close to its flatband condition using model parameters extracted from the DFT calculations.  The electronic structure is most interesting in the AdA case, on which we subsequently focus, because its flat bands emerge cleanly and are accompanied by strongly dispersive bands covering the same energy interval.

\ifpdf
\begin{figure}[htp]
    \includegraphics[width=1.\linewidth]{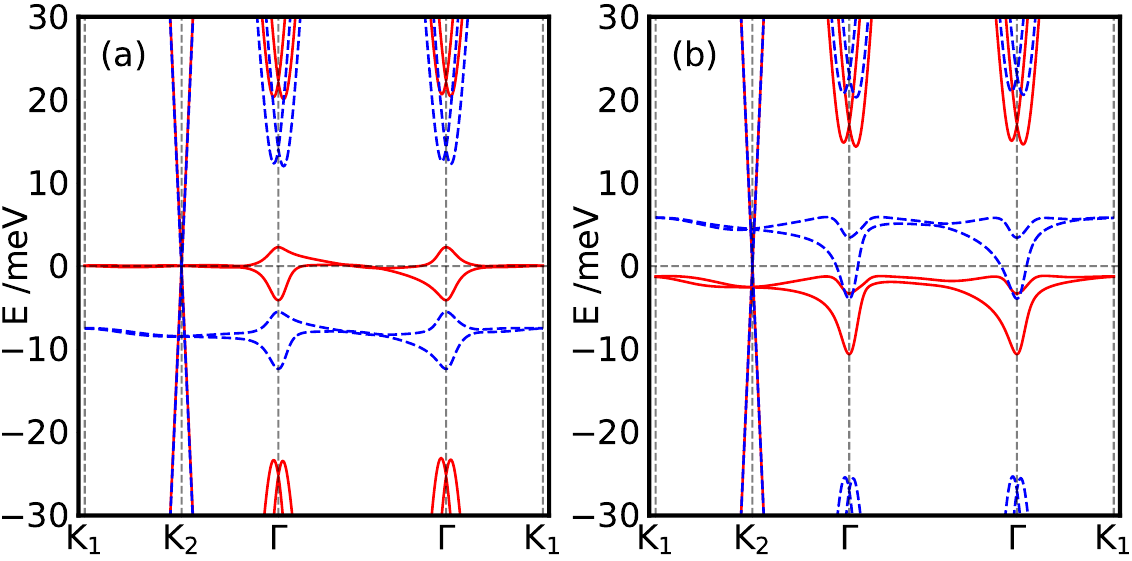}
    \caption{Moir\'e bands of TTLG in the AdA stacking configuration at a twist angle of 1.55$^{\circ}$, successively adding terms from the continuum model in Eq.~\ref{equ:phenomen}. (a) Bands calculated including only the twisted layer tunneling ($\mathcal{T}$) and intra-layer Dirac cone ($H_{D}$) terms in the Hamiltonian (red) and bands when the site energy term ($H_{s}$) is also included (dashed blue). (b) Bands when hopping between different sites of aligned layers are also included ($H_{t}$). The red and blue curves respectively include and neglect coupling between the outer layers.
    }
    \label{fig:flatbands}
\end{figure}
\fi

To demonstrate the relevance of the smaller model-parameters in the continuum model Hamiltonian that we have estimated using DFT, we point to the band alignment of AdA stacked TTLG. We compare moir\'e band calculations based either on only the dominant $\mathcal{H}_D$ and $\mathcal{T}$ terms in the model Hamiltonian (see Eq.~\ref{equ:phenomen}) or also on the on-site energies $\mathcal{H}_s$ (see Fig.~\ref{fig:flatbands})). We find that the effect of $H_{s}$ is to induce an energy difference between the Dirac point of the dispersive bands and the flat band energy.  Intralayer corrections to the Fermi velocity (the $t$'s from Table~\ref{tbl:intralayer_coupling_elements}) and direct hopping between outer layers (the $t$'s from Table~\ref{tbl:aligned_coupling_elements}) further modulate the bands in a non-trivial manner (see Fig.~\ref{fig:flatbands}) , particularly by changing the shape and relative alignment of the flat bands. Indeed, all of these contributions to the Fermi alignment are of the same order of magnitude. A quantitative description of the final alignment between the highly dispersive bands and the flat bands therefore requires carefully considering all of these terms.

\ifpdf
\begin{figure}[!htp]
\includegraphics[width=0.9\linewidth]{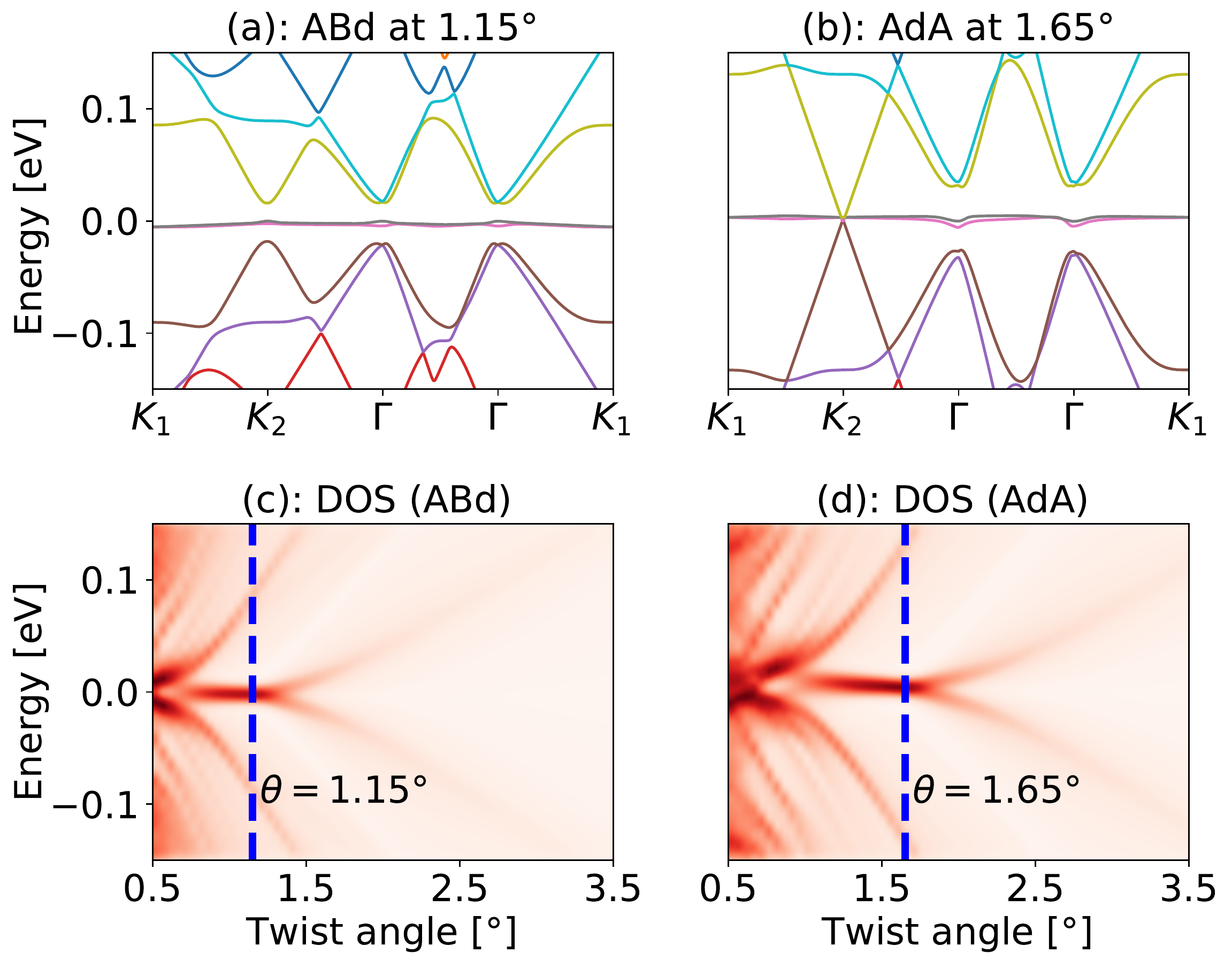}
\caption{
    Simplified bandstructure and density of states (DOS) of ABd (a,c) and AdA (b,d) stacked TTLG. (a) and (b) are the bandstructures of top-layer (ABd) and middle-layer (AdA) twisted trilayers at a particular close-to-magic twist-angle, and (c) and (d) are twist-angle-dependent DOS with the twist angles of (a) and (b) labeled by blue dashed lines. }
    \label{bandstructure}
\end{figure}
\fi

While all small contributions in our ab-initio model are relevant for the detailed analysis of the band evolution, simplified models suffice for discussing qualitative band structure properties. Importantly, the symmetries that dictate many band properties, such as the mirror symmetry for the AdA case, are present in both the full and the simplified model. In the following we use the simplified version, including only $\mathcal{H}_D$, $\mathcal{H}_s$ and $\mathcal{T}$; for ABd stacking we additionally use the values from Table~\ref{tbl:aligned_coupling_elements} in the Appendix. The simplified models facilitate the comparison of AdA and ABd electronic structure in Fig.~\ref{bandstructure}, and predict the difference of the first magic angles for top-layer (ABd) and middle-layer (AdA) twisted trilayers, {\it i.e.} first magic angle for the AdA stacking is $\approx \sqrt{2}$ times larger ABd case.

AdA stacked TTLG has a mirror symmetry ($\hat{z} \rightarrow -\hat{z}$), which leads to even and odd parity bands and allows the even-parity flat and odd-parity dispersive bands to cross \cite{Khalaf2019,Mora2019}. As we will show in the following, this mirror symmetry is important, especially for the density of states (DOS) and for transport properties. The decoupled flatbands and dispersive Dirac bands hybridize whenever the mirror symmetry is broken, for example by substrate effects, by applying an external electric field or by shifting the top layer laterally \cite{Khalaf2019}. With broken mirror symmetry the width of flatbands, and therefore electronic correlation strengths, can be tuned by applying an electric field.

\subsection{Experimentally Relevant Band Characteristics}\label{band_characteristics}

We now focus on AdA single-layer-twist trilayers.  In order to provide a sense for the experimental relevance of the electronic structure variability that we analyze, we provide results not only for the energy band dispersion and density-of-states, but also for other experimentally relevant band-structure characteristics - {\it i.e.} other observables that depend only on the energy bands. 

\subsubsection{Drude Weight}

The Drude weight (also known as the charge stiffness) measures the inertia of the electronic response to external electric fields.  In the case of metallic bands, we can utilize Kohn's formula\cite{Kohn1964} to calculate the Drude weight from band velocities $v_{n\bk\mu} = \langle n\bk |\partial H/ \partial k_{\mu} |n\bk \rangle/\hbar $, via:
\begin{equation}\label{drude_formula}
    D_{\mu\nu} = \frac{e^2}{2\pi} \sum_{n}\int_{BZ}d\bk \,
    \frac{\partial f_{n\bk}}{\partial \varepsilon} \; v_{n\bk\mu} v_{n\bk\nu}
\end{equation}
In the case of interest, $C_{3v}$ symmetry in a two-dimensional electron system ensures that the Drude weight tensor is proportional to the unit matrix \cite{Malgrange2014}, {\it i.e.} $D_{xx} = D_{yy}$ and $D_{xy}=D_{yx}=0$, as detailed shown in App.~\ref{drude}.  The longitudinal conductivity is proportional to the product of the Drude weight and the transport scattering time.

\subsubsection{Weak-field Hall Conductivity}

Using the Jones-Zener solution of the Boltzmann equation \cite{Hall1991},
\begin{equation}
  \sigma_{xy} = 2\frac{e^3}{\hbar}B \int d\bk
    \frac{-\partial f_{\bk}}{\partial \varepsilon} (v_y \tau_{\bk}) (v_y\partial_{k_x} - v_x \partial_{k_y})(v_x \tau_{\bk} ), 
\end{equation}
we also calculate the weak-field Hall conductivity $\sigma_{xy}$ as a function of displacement field and carrier density (see Fig.~\ref{drude_hall_field} (d)). Here $\tau_{\bk}$ is the relaxation time, which we approximate $\tau_{\bk}\approx \tau$ by a constant, and $ \partial_{k_{\mu}} \equiv \partial/\partial k_{\mu} $. We express the Hall conductivity in units of $(e^2/h) (2[eV]\tau/\hbar \cdot a/l_{B})^2/\pi$ where $l_{B}=\sqrt{\hbar/eB}$ is the magnetic length and $a$ is the lattice constant of graphene. At weak magnetic fields the Hall resistivity $\rho_{xy}=-\sigma_{xy}/\sigma_{xx}^2$ is independent of the transport scattering time and is a pure band-structure property, and in the case of isotropic bands is inversely related to the carrier density.  

\subsubsection{Carrier Density and Hall Density}

\ifpdf
\begin{figure*}[htp]
\includegraphics[width=0.8\linewidth]{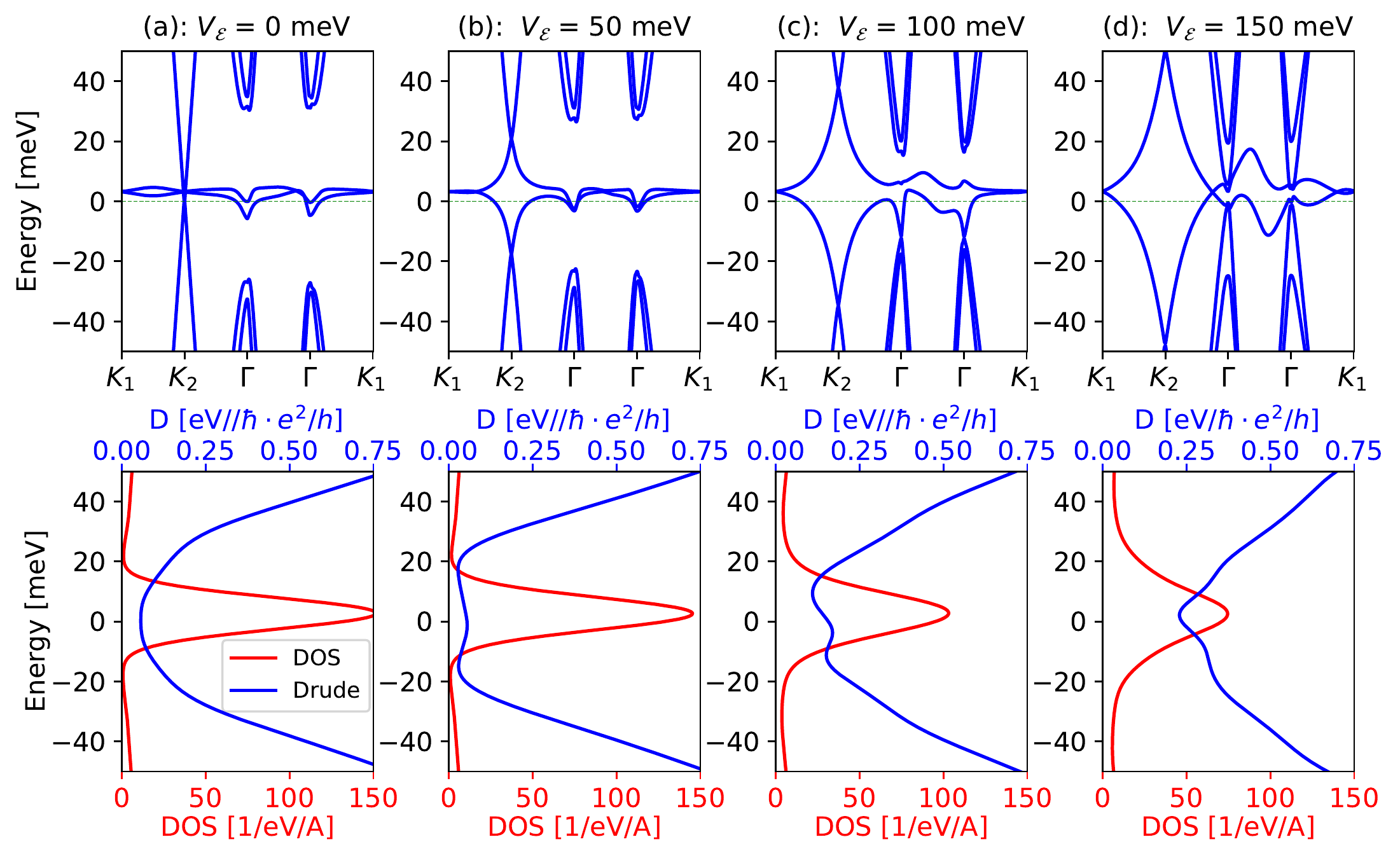}
    \caption{Bandstructure and transport properties of AdA stacked TTLG for various displacement fields. The top panels of (a)-(d) show bandstructures as the displacement potential increases from 0 to 150 meV.  Mirror symmetry breaking couples the even parity flat bands to the odd-parity dispersive Dirac cone bands.  The Dirac points of the outer layers are pushed away from the Fermi level, but are protected by symmetry. The lower panels of (a)-(d) illustrate the evolution of the DOS and the Drude weight with displacement potential.}
    \label{flatband_field}
\end{figure*}
\fi

To facilitate comparison with experimental studies of samples with gate controlled carrier densities, we will sometimes express our results in terms of carrier density instead of Fermi energy. The mapping from chemical potential to carrier density can be obtained by integrating the DOS $D(E)$, 
\begin{equation}
    n = 4 \cdot \frac{1}{\Omega_m} \int_{E_0}^{E_F} D(E) f(E)dE, 
\end{equation}
where $ \Omega_m = |\bR_1|^2 \sin{\pi/3} $ is the area of a moir\'e unit cell and $ |\bR_1| \approx |\ba_1|/\theta$ is the length of the moir\'e period. The factor 4 accounts for valley and spin degeneracy. $E_0$ is the Fermi energy at neutrality, where the bands of the moir\'e system are half filled, and $f(E)$ is the Fermi-Dirac distribution function.  The carrier density can be compared directly with the so-called Hall density,
\begin{equation} 
    n_H \equiv \frac{B}{\rho_{xy}ec}=\frac{B \tau^2 D_{xx}^2 }{\sigma_{xy}ec},
    \label{eq:Hall_density}
\end{equation}
which is independent of transport lifetime and therefore a band characteristic. The expressions quoted above apply only in the relaxation time approximation. In the limit of a single isotropic closed Fermi surface, the Hall density is equal to the carrier density.

\subsection{Mirror Symmetry and Displacement Fields}\label{section_electric_field}

We now discuss mirror symmetry breaking by gate-controlled displacement fields.  In our calculations the influence of the displacement field is modelled by adding a term to the Hamiltonian that creates a relative difference between the energies of sites located in different layers:
\begin{equation}
    H_{\mathcal{E}}(\br) = \sum_{l,\alpha}V_l(\mathcal{E}) 
    c_{l\alpha}^{\dagger}(\br) c_{l\alpha}(\br).
\end{equation}
In the following we will set the displacement potential in the middle layer to $0$, and in the outer layers to $\pm V_{\mathcal{E}}$.

In the top panels of Fig.~\ref{flatband_field} we illustrate the dependence of the flat bands around the Fermi energy on displacement potentials.  We see that hybridization between the odd parity dispersive bands and the even parity flat bands increases steadily with displacement field, with the result that all bands become dispersive.  The $C_2T$ symmetry that allows Dirac points at the moir\'e Brillouin-zone corner is not broken by applied displacement fields and linear band crossings therefore remain - although they are shifted in energy. We see in the top panels of Fig.~\ref{flatband_field} that the total width of the broadened flatbands is approximately proportional to the displacement potential. In the bottom panels of Fig.~\ref{flatband_field} we show that the DOS (red curves) peak is close to the neutrality point at all displacement fields.  By contrast the Drude weight has a minimum in the vicinity of the neutrality point, due to the flatness (very low band velocities) of these subbands (see lower panel of Fig.~\ref{flatband_field} (a)).  When a finite displacement field is present, the Dirac points originating from the odd-parity bands and the minima of the Drude weight are energetically shifted.  Local minima of the Drude weight emerge at the Dirac points and are present until these touch the edges of the surrounding bands (See lower panel of Fig.~\ref{flatband_field}). Further increases in the displacement field move the minimum of the Drude weight back to neutrality. Two trends are at play: (i) the width of the flatbands increases, raising the band velocity. The Drude weight therefore increases despite the decrease in the DOS. (ii) the Dirac points (where the DOS is minimal) of the outer layers are shifted away from the charge neutrality point $E=0$. The Drude weight is then mainly determined by the band velocities, which become minimal near $E=0$ (lower panel of Fig.~\ref{flatband_field}).

\ifpdf
\begin{figure}[htp]
\includegraphics[width=1.\linewidth]{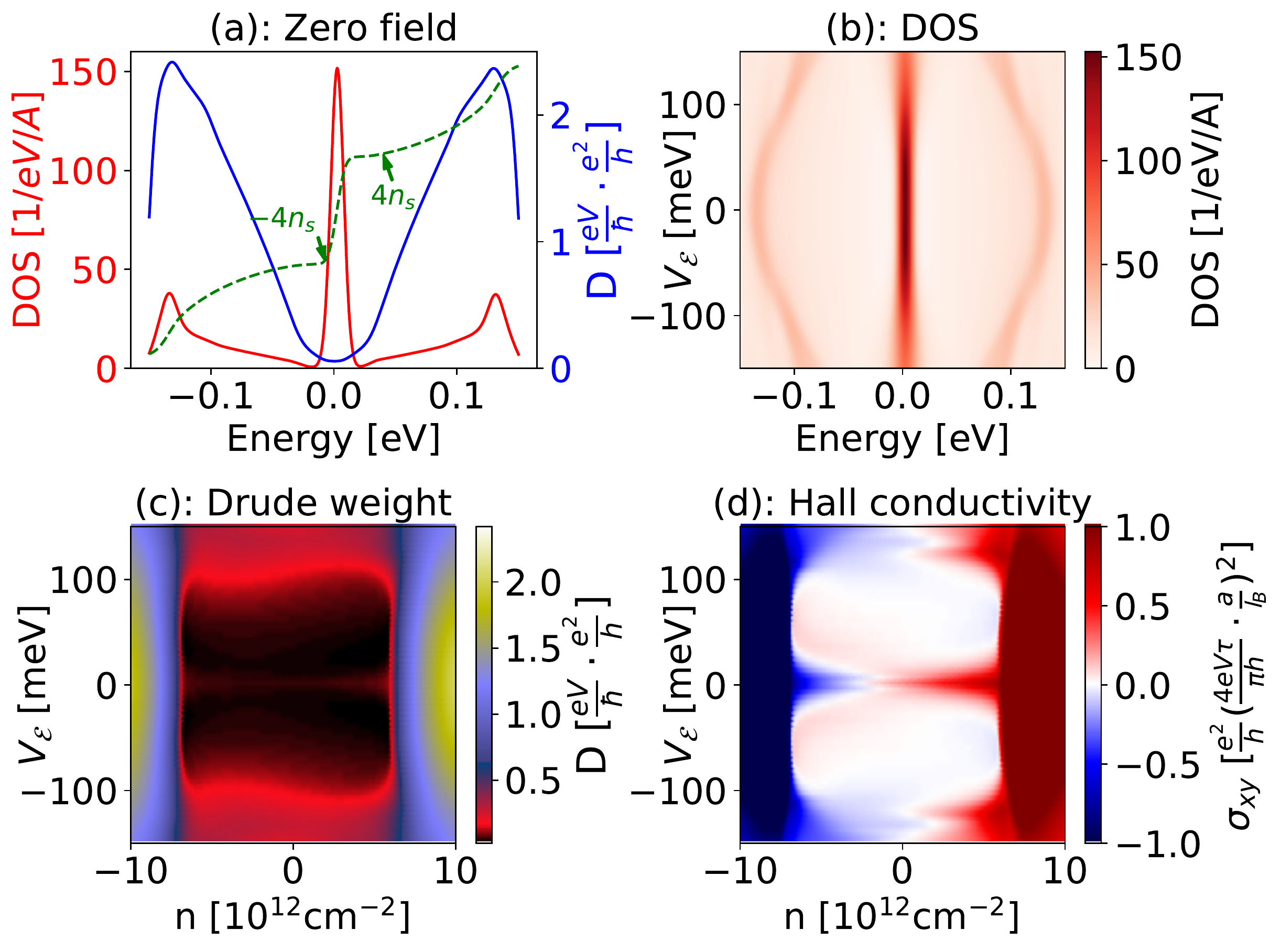}
    \caption{Drude weight (in units of $eV/\hbar \cdot (e^2/h)$) and Hall conductivity as a function of carrier density and displacement potential. (a) The lower panel of Fig.~\ref{flatband_field}(a) plotted over a wider range of Fermi energy.  The dashed green line is a plot of the carrier density. The scale of the carrier density plot is implied by the green arrows that mark where $n=\pm 4n_s$, with $n_s$ one electron per moir\'e unit cell ($n_s\approx 1.58\times10^{12} cm^{-2}$). (b) Density of states {\it vs.} Fermi level with the displacement potential varied from -150 meV to 150 meV. (c) Drude weight ($2D = D_{xx} + D_{yy}$) and (d) Hall conductivity ($\sigma_{xy}$), plotted as a function of displacement field and carrier density.}
    \label{drude_hall_field}
\end{figure}
\fi

\ifpdf
\begin{figure}[htp]
\includegraphics[width=1.\linewidth]{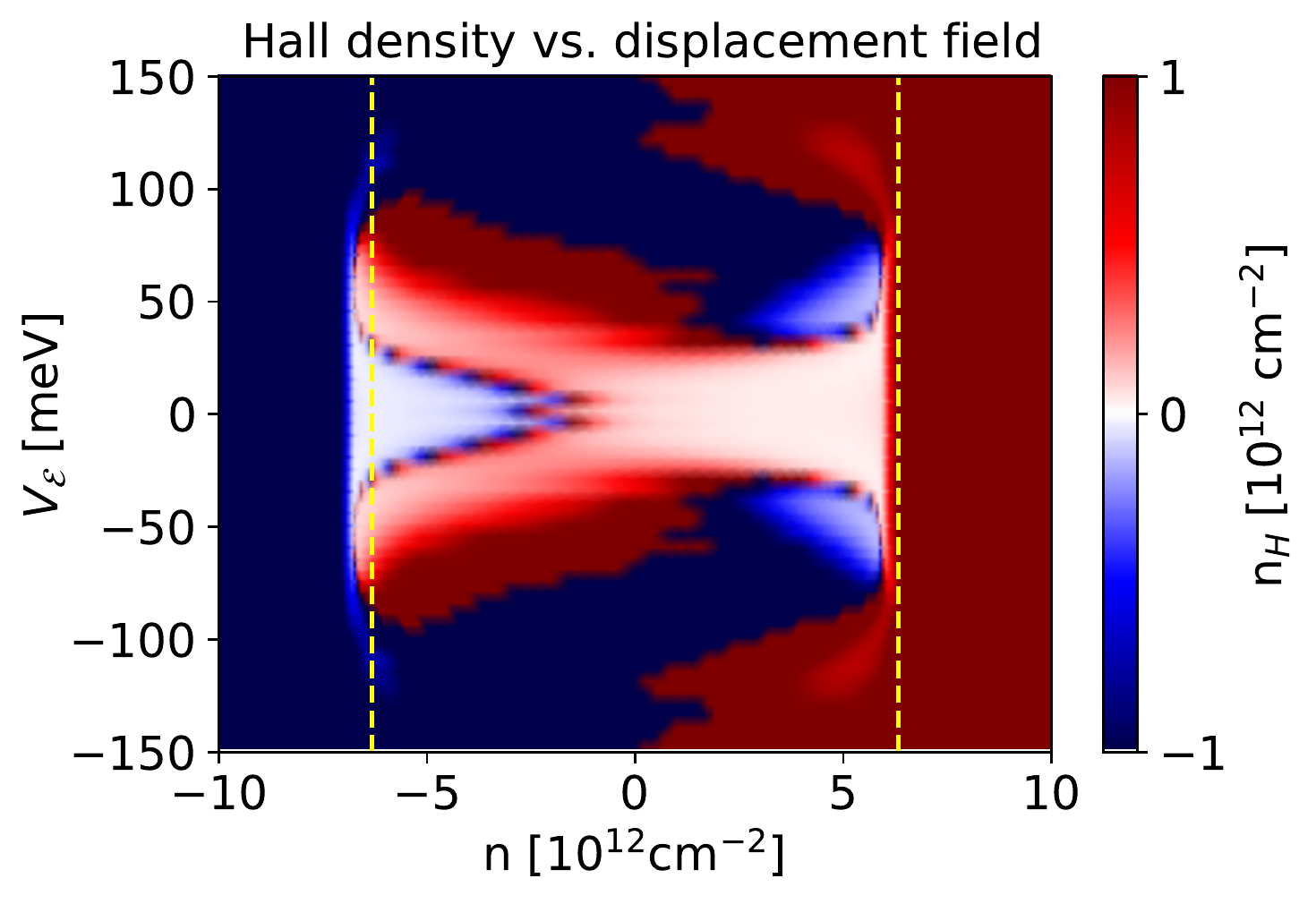}
    \caption{Hall density ($n_H$) as a function of carrier density ($n$) and displacement potential ($V_{\varepsilon}$).  The dashed yellow lines mark the densities $\pm 4 n_s$. Between these densities the Fermi level lies within the flat bands.}
    \label{fig:drude_hall_field_density}
\end{figure}
\fi

The Drude weight's dependence on Fermi energy in Fig.~\ref{drude_hall_field} (a) is converted in Fig.~\ref{drude_hall_field} (c) to a dependence on carrier density. The carrier density ($n$) changes rapidly with Fermi energy in the region of the flatbands (see Fig.~\ref{drude_hall_field} (a)).  We find that the maximum of the DOS is pinned to the flat band Dirac point (compare Fig.~\ref{drude_hall_field} (a)). The edges of the subbands closest to the flatbands in energy tend to move together with increasing displacement potential.  The Drude weight (Fig.~\ref{drude_hall_field} (c)) is small over a wide energy range near neutrality, as expected given the ultra-low band velocities discussed above.

Analyzing the resulting Hall conductivity ($\sigma_{xy}$) we find that --- for filled flat bands --- the carriers have electron character even below neutrality, highlighting that the dispersive Dirac bands play a dominant role in the Hall conductivity around half-filling, and that the Dirac point of the dispersive band lies below the flat bands in energy.  Furthermore, the electron character of the Hall conductivity in general only weakly depends on the displacement potential. In Fig.~\ref{fig:drude_hall_field_density} we compare the actual carrier density with the Hall density, defined in Eq.~\ref{eq:Hall_density}.  Comparisons of Hall density and total carrier density are readily made experimentally and are therefore a convenient point of comparison with theory. We see in Fig.~\ref{fig:drude_hall_field_density} that the Hall density is strongly dependent on displacement field when the Fermi level lies within the flat bands, and much more weakly so when the Fermi level lies outside the flat bands.

\subsection{Mirror Symmetry and Lateral Stacking Shifts}
\label{section_stacking_shift}

\ifpdf
\begin{figure*}[htp]
\includegraphics[width=0.8\linewidth]{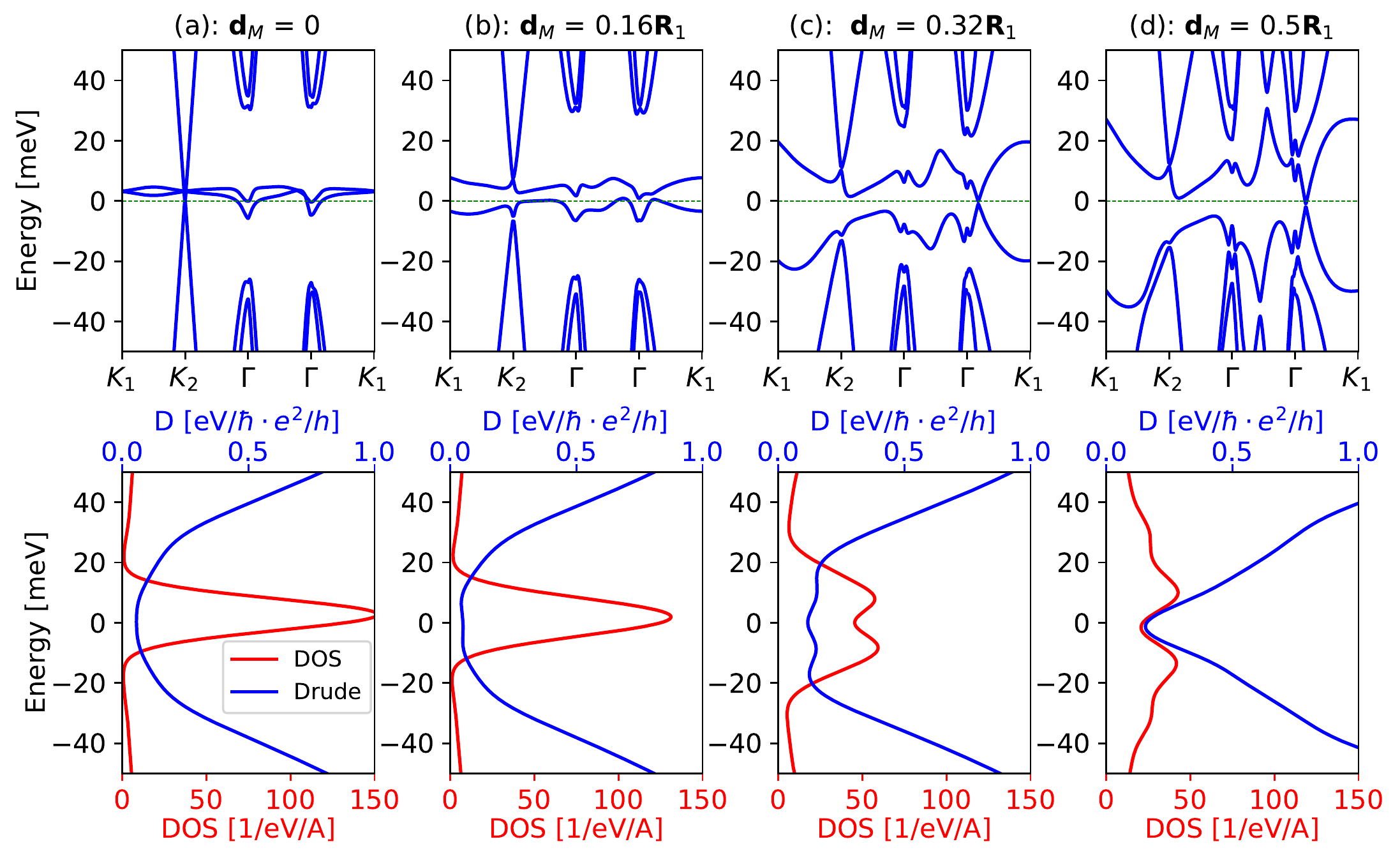}
    \caption{Band and transport properties of TTLG with a twisted middle layer for several lateral stack shifts along the $\bR_1$ direction.  (a)-(d) (top panels) bandstructures,  (lower panels) DOS and Drude weight: Shifting $\bd_M$ away from zero breaks the mirror symmetry and the $C_{2}T$ symmetry, causing the Dirac points to gap.}
    \label{flatband_dis}
\end{figure*}
\fi

Because accidental lateral stacking shifts are present in all current trilayer devices, it is important to understand their influence on electronic properties. In contrast to the trilayer case, lateral stacking shifts have no influence on the electronic structure of bilayers \cite{Bistritzer2011,Khalaf2019}. For the TBLG case, the interlayer tunneling Hamiltonian in Eq.~\ref{equ:tunneling_amp} is,  
\begin{equation}\label{tunnel:bilayer}
    \mathcal{T}(\br) = \sum_{\alpha\beta^{\prime};n}
    w^{BL}_{\alpha\beta^{\prime}} \;
    e^{i\bG_n\cdot (\bt_{\alpha} - \bt_{\beta^{\prime}} ) }
    e^{-i\bq_n \cdot \br} c_{\alpha}^{\dagger}(\br) c_{\beta^{\prime}}(\br),
\end{equation}
where the bilayer interlayer tunneling $w^{BL}_{\alpha\beta^{\prime}}$ has no layer-dependent index and we explicitly denote the phase shift introduced by the atomic positions $\bt_{\alpha(\beta^{\prime})}$. A general change in layer alignment at the origin ($\tau_{\beta^{\prime}} \rightarrow \tau_{\beta^{\prime}} + \bd_{trans}$) can be expressed as a lateral stacking shift $\bd(\br_{trans}) = \theta \hat{\bm z} \times \br_{trans}$ and thus simply manifests as a extra phase $e^{i \tilde{\bG}_n \cdot \bd(\br + \br_{trans})}$ in Eq.~\ref{tunnel:bilayer}, without changing the phase of the interlayer tunneling matrix elements $\omega_{\alpha\beta'}^{BL}$. Translating the layer that is rotated relative to the two other layers (e.g. shifting the top layer in a ABd configuration) in a TTLG therefore leaves the electronic structure - in analogy to the TBLG case - unchanged. In contrast, changing the relative lateral stacking of the two aligned layer strongly alters the electronic structure of the TTLG.  Since all configurations with a twisted outer layer except the ABd configuration depend on accidental lateral displacements, we herein focus on the case of middle-layer twisted trilayers.

Translating the moir\'e patterns involving the top layer changes the value of all coupling terms, as highlighted, {\it e.g.}, by the differences between AdB and AdA stacking in Tables~\ref{tbl:coupling_elements},~\ref{tbl:intralayer_coupling_elements}, and ~\ref{tbl:aligned_coupling_elements}. However, these changes are small and we will neglect them. The remaining moir\'e shifts can be accounted for entirely by phase factor changes\cite{Khalaf2019} for tunneling between the top layer and the other two layers, {\it i.e.} in $w^{b,m,\bG_n}_{\alpha\beta} $ and $w^{m,t,\bG_n}_{\alpha\beta}$. We write a general spatial shift of the moir\'e pattern as 
\begin{equation}
    \bd_M = \lambda_1 \mathbf{R}_1 + \lambda_2 \mathbf{R}_2, 
\end{equation}
where $ \mathbf{R}_1 $ and $ \mathbf{R}_2 $ are the lattice vectors of the moir\'e unit cell. Since properties can depend only on the displacement modulo moir\'e lattice vector shifts, we can restrict our attention to $ 0 \le \lambda_1, \lambda_2 \le 1 $. These moir\'e pattern shifts correspond to top layer displacements by $ \bm{\lambda} = \theta \hat{z} \times \bd_M = \lambda_1 \mathbf{a}_1 + \lambda_2 \mathbf{a}_2$ with $\mathbf{a}_1/\mathbf{a}_2$ the lattice vectors of graphene.  The phase factors that account for lateral displacements are therefore $\exp(i\mathbf{G}_n \cdot \bm{\lambda} )$, where $\mathbf{G}_n \cdot \bm{\lambda} = \lambda_1 \bG_n \cdot \mathbf{a}_1 + \lambda_2 \bG_n \cdot \mathbf{a}_2$.

The top layer lateral stacking shift breaks not only the mirror symmetry but, unlike an applied external displacement field, also $C_{2}T$ symmetry. As illustrated in the top panel of Fig.~\ref{flatband_dis}, the flat-band Dirac cone is therefore gapped when $\bd_M \ne 0$, and the dispersive band Dirac points are both gapped and, because of mirror symmetry breaking, shifted away from the flat bands. When the lateral stacking shift moves from $\bd_M=0$ along $\mathbf{R}_1$, the maximum in the Fermi level DOS shifts away from neutrality. Lateral stacking shifts split the flat band DOS peak , and the Drude weight develops two local maxima around its global minimum (lower panels of Fig.~\ref{flatband_dis}).

\ifpdf
\begin{figure}[htp]
    \includegraphics[width=0.98\linewidth]{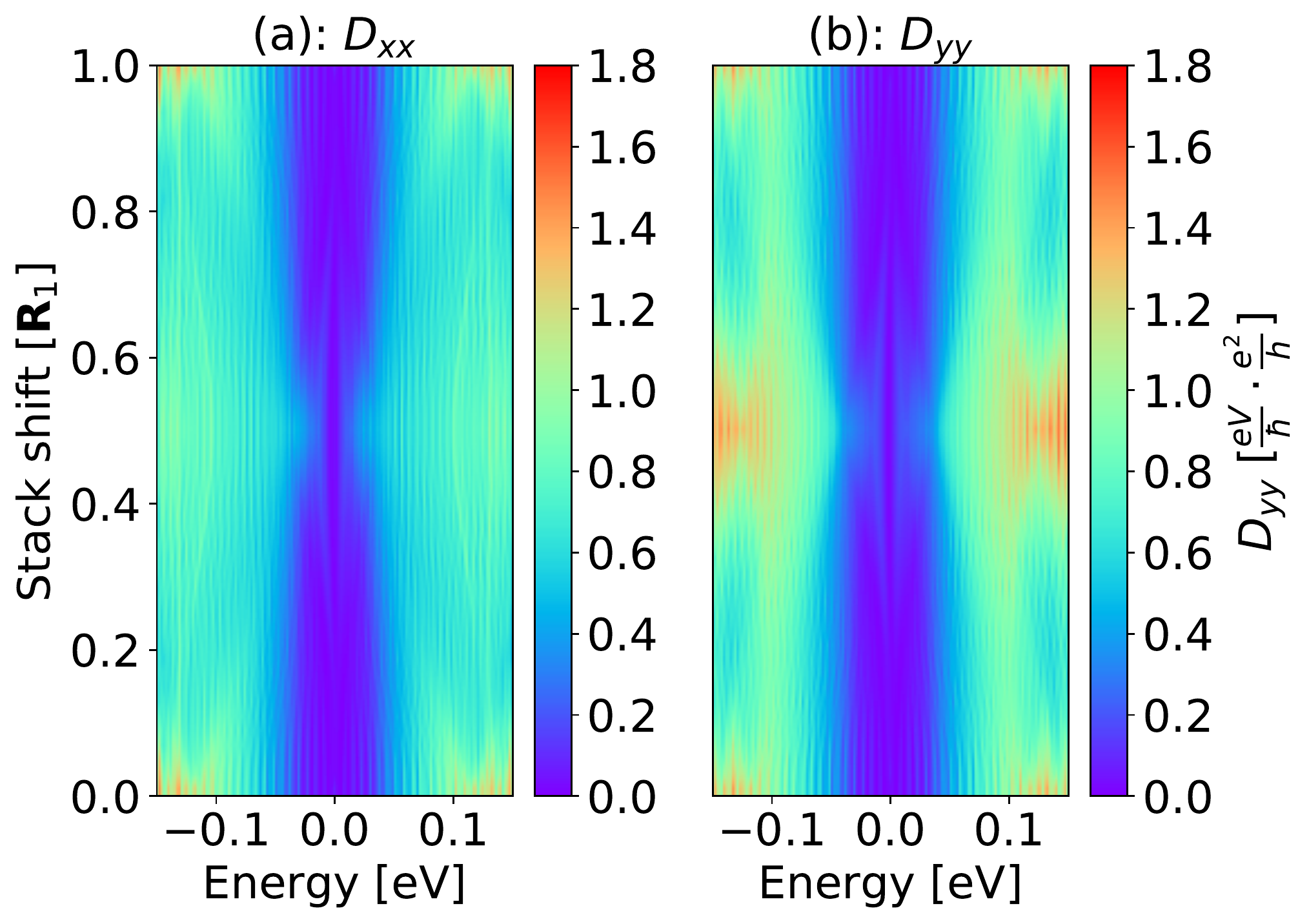}
    \caption{Drude weight $D_{xx}$ and $D_{yy}$ versus the Fermi energy. Unlike the case where we included a displacement field, lateral stacking shifts break the $C_{3v}$ symmetry and lead to anisotropic transport. }
\label{shift_dir}
\end{figure}
\fi

For $\bd_M = 0$, the Hamiltonian features $C_{3v}$ symmetry for all displacement fields guaranteeing isotropic linear-response transport coefficients (see App.~\ref{drude}). By contrast, $C_{3v}$ symmetry is lost when $\bd_M \ne 0$.  To quantify the degree of transport anisotropy produced by accidental lateral stacking shifts we consider $D_{xx}$ and $D_{yy}$ as a function of Fermi energy and $\bd_M$ along $\bR_1$ (see Fig.~\ref{shift_dir} (a) and (b)).  The difference between $D_{xx}$ and $D_{yy}$ becomes maximal at $\bd_M = \bR_1/2$ (Compare Fig.~\ref{shift_dir} (a) and (b)). It follows that measuring the transport anisotropy can be extremely valuable in characterizing twisted trilayer devices.

In Fig.~\ref{stack_shift}) we examine the influence of lateral stacking shifts on the flatband width, the Drude weight, and the Hall conductivity.  For the perfect AdA stacking configuration without a lateral stacking shift, the Drude weight ($D = D_{xx} + D_{yy}$) is always small for carrier densities in the flatband range $ -4n_s \le n \le 4n_s $ (See Fig.~\ref{stack_shift} (a)). In this range of density the Hall conductivity varies approximately linearly with carrier density and features an $n$-type sign at charge neutrality.

\ifpdf
\begin{figure}[htp]
\includegraphics[width=0.9\linewidth]{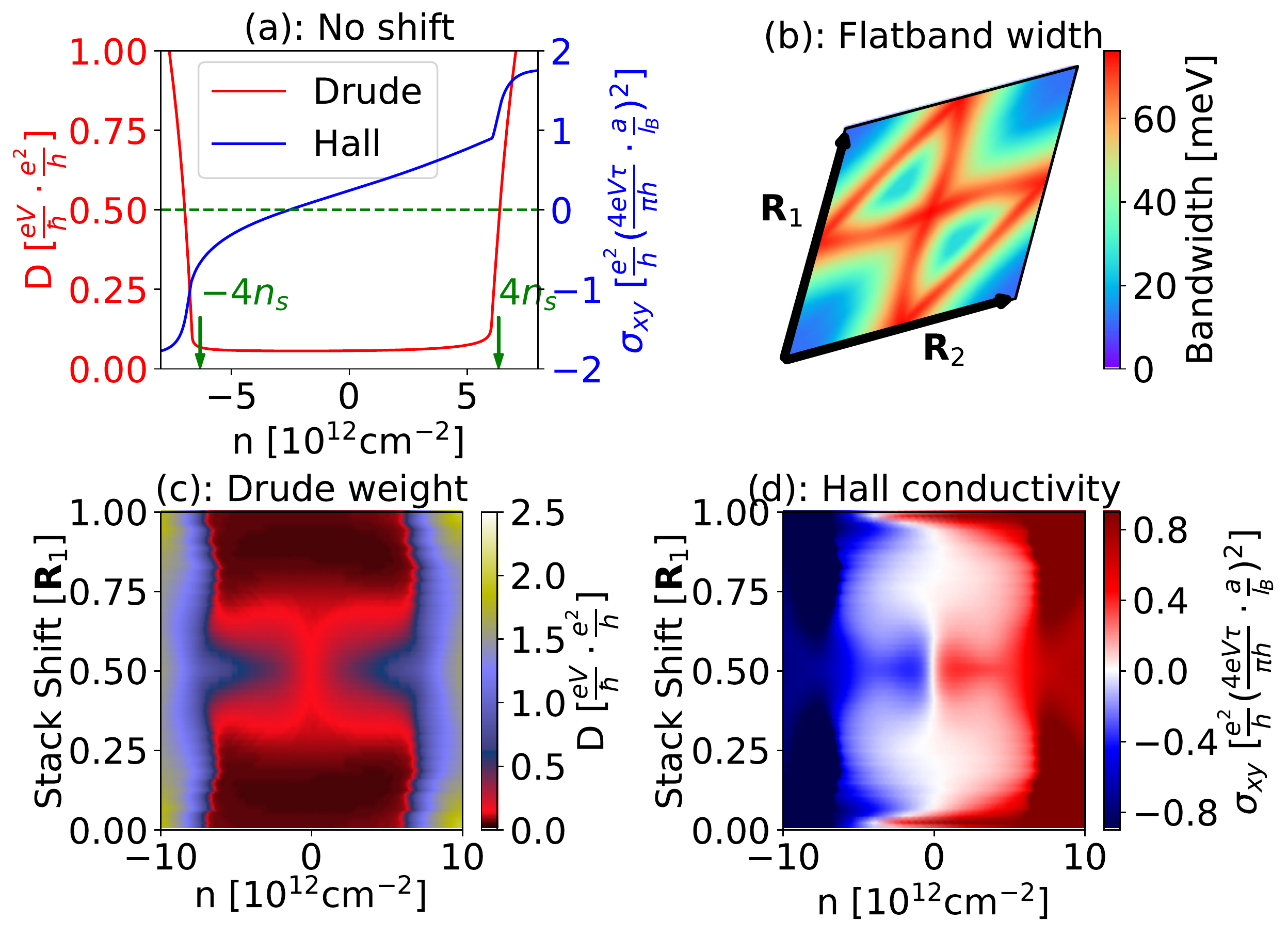}
    \caption{Influence of stacking-shift on experimentally accessible observables. (a) Drude weight ($D = D_{xx} + D_{yy}$) and Hall conductivity ($\sigma_{xy}$) at zero lateral stacking shift ($\bd_M=0)$. (b) Flatband width, defined as the difference in energy between the top of the first band above neutrality and the bottom of the first band below neutrality, {\it vs.} top layer stacking shifts over one moir\'e unit cell bounded by $\bR_1$ and $\bR_2$ (see also Fig.~\ref{gaps} (a)). The bandwidth, which is minimal at zero lateral stacking shift, reaches up to around 70meV for some configurations. (c) Drude weight and (d) Hall conductivity {\it vs.} $\bd_M$ along the $\bR_1$ direction.}
\label{stack_shift}
\end{figure}
\fi

The evolution of the width of the flat bands as a function of stacking shift (Fig.~\ref{stack_shift} (b)) reveals that the two bands closest to charge neutrality are not narrow at all lateral stacking shifts, reaching up to $~70$ meV - close to typical bandwidths in graphene multilayer moir\'es \cite{Bistritzer2011,Kim_2017}. The flatband width is minimal near $ \bd_M = 0 $ and close to the $\bd_M = (\bR_1 + 2\bR_2)/3$ and $\bd_M = (2\bR_1 + \bR_2)/3$ lines. Magic angle behavior is therefore common but not universal as a function of lateral stacking shift.  Correspondingly, the Drude weight near neutrality increases when the top layer is shifted away from the AdA configuration. Generally speaking, it still remains small compared to values outside the flatband region of carrier density.  The increase in Drude weight is linked to the increase of band velocity due to increased flatband width. The change in sign of the Hall conductivity (see Fig.~\ref{stack_shift} (d)) shifts towards neutrality for non-zero $\bd_M$. The evolution of the Hall densities with $n$ at different stacking shifts (Fig.~\ref{fig:drude_hall_dis_density}) shows a similar trend: for $\bd_M=0$, the Hall density changes sign not at neutrality but at a negative carrier density. The reason is a competition between the contributions from the odd-parity dispersive band and the even-parity flatband to the Hall conductivity. When the two sets of bands are coupled by $\bd_M \ne 0$, the Hall density sign change generally occurs much closer to charge neutrality.  Measurements of the Hall density can therefore be used to determine the strength of mirror symmetry breaking in AdA trilayers.

\ifpdf
\begin{figure}[htp]
\includegraphics[width=1.\linewidth]{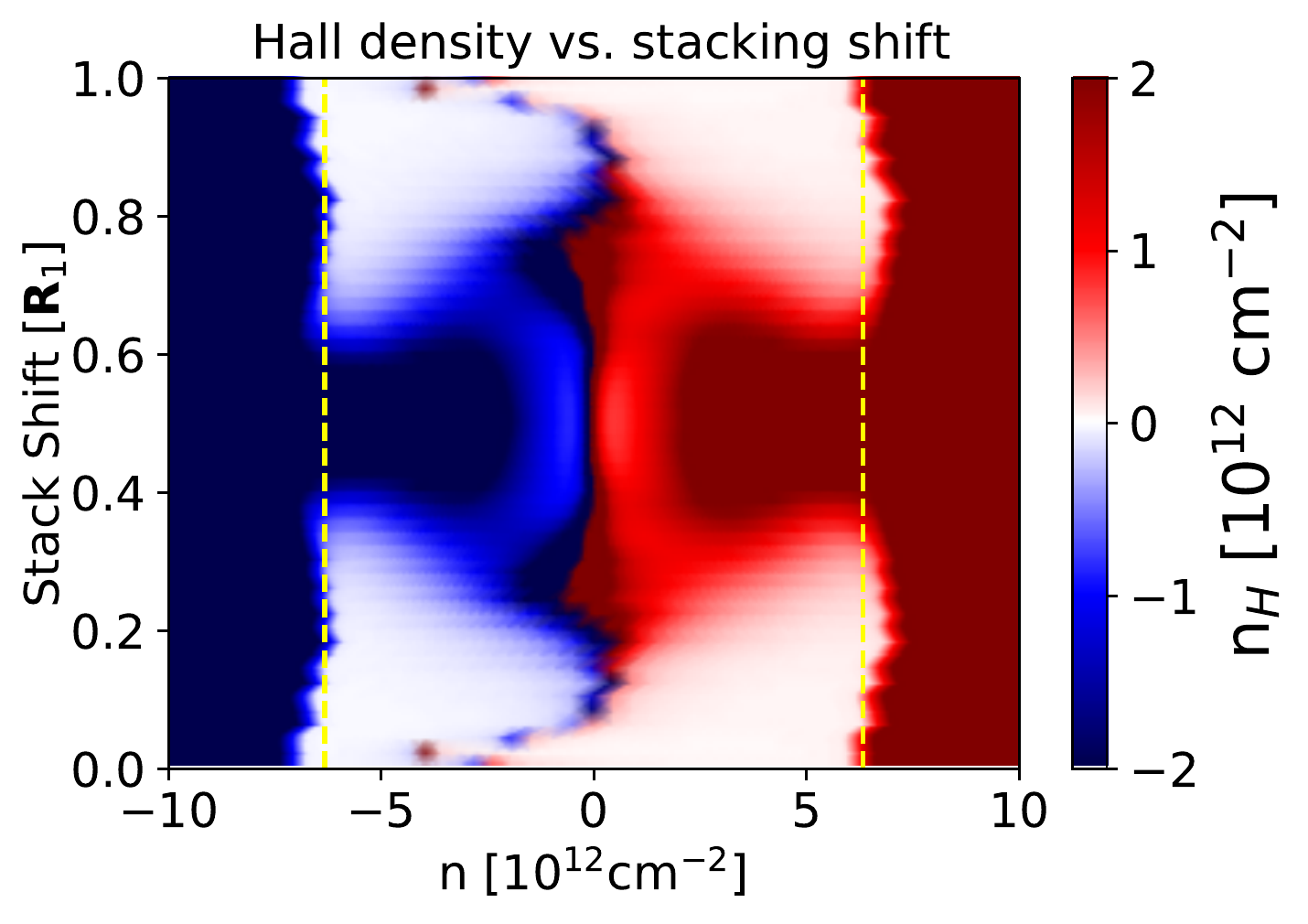}
    \caption{Hall density ($n_H$) as a function of carrier density ($n$) and lateral stacking shift ($\bd_M$). The Hall density at neutrality is n-type for $\bd_M = \mathbf{0}$, and much closer to $0$ for $\bd_M \ne \mathbf{0}$. Hall density measurements are sensitive to the strength of coupling between even and odd parity bands.}
    \label{fig:drude_hall_dis_density}
\end{figure}
\fi

\ifpdf
\begin{figure}[htp]
\includegraphics[width=0.9\linewidth]{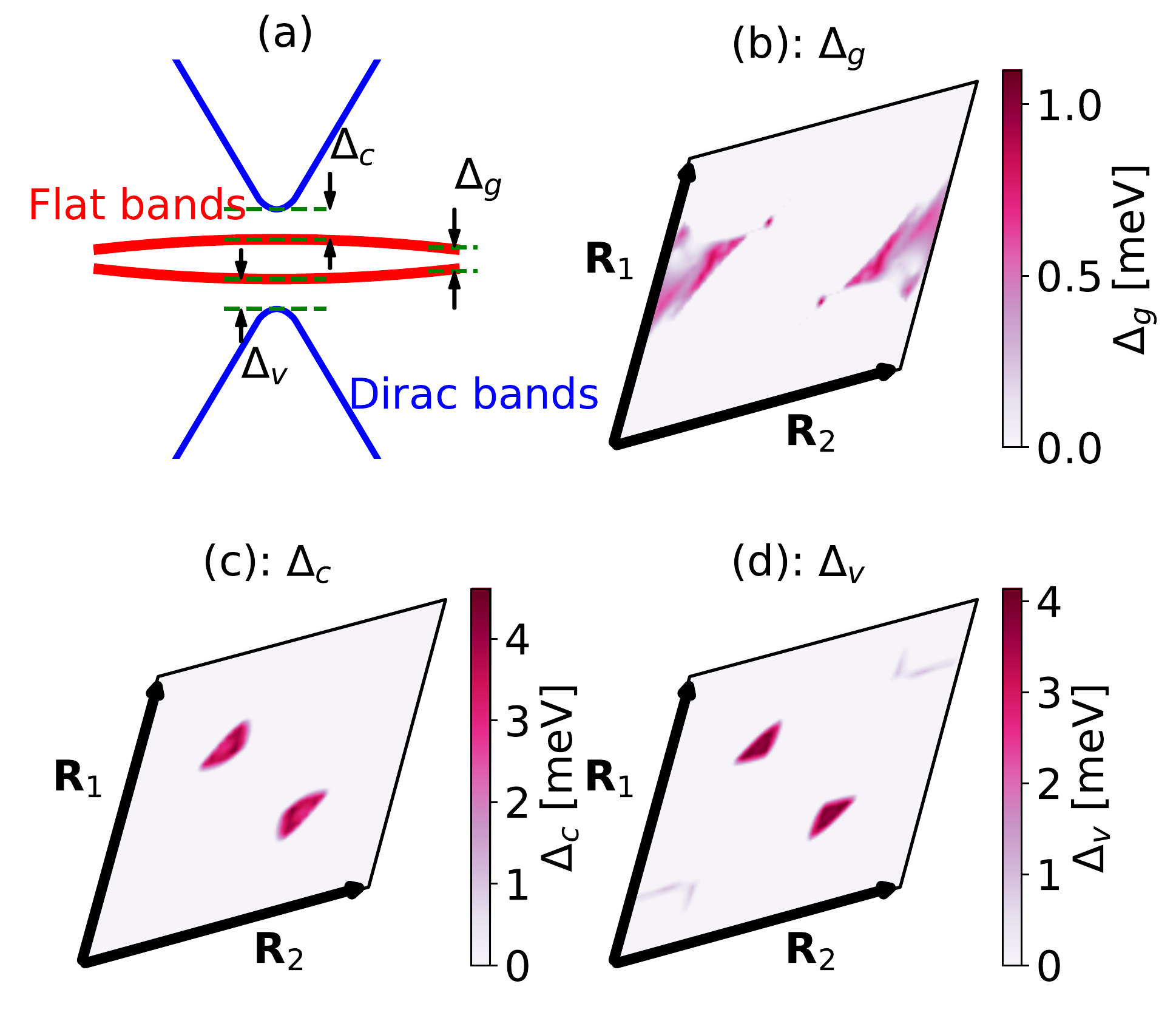}
\caption{
    Evolution of the gaps between AdA minibands.  The band gaps $\Delta_g$, $\Delta_c$ and $\Delta_v$ are defined in (a). The dependence of their magnitudes on lateral stacking shift is plotted in (b)-(d). For most stacking configurations no gap is present. However for some stacking configurations a small gap opens between the flatbands shown in (b), as well as between the flat bands and the dispersive bands shown in (c) and (d).
}
\label{gaps}
\end{figure}
\fi

Finally, we comment on the band gaps that are opened by coupling between even and odd parity bands at $\bd_M \ne 0$.  We identify three different gaps: $\Delta_g$ between the flat conduction band (first band above neutrality) and the flat valence band (first band below neutrality), $\Delta_c$ between the first and second conduction bands, and $\Delta_v$ between the first and second valence bands (see Fig.~\ref{gaps}~(a)). In Fig.~\ref{gaps}~(b)-(d) we show that gaps are present for some lateral stacking shifts in our single particle picture. There is, however, no shift for which $\Delta_g \ne 0$ and $\Delta_{c}/\Delta_v \ne 0$ simultaneously, highlighting the rarity of isolated gapped flatbands in a TTLG system. 

\section{Summary and Discussion}\label{section_conclusionss}

We have used density-functional-theory input to construct a continuum model for twisted trilayer graphene that allows for moir\'e-induced spatial variation not-only in tunneling between adjacent layers but also in hopping between top and bottom layers, in Fermi velocity, and also in the site-energies of both interior and exterior layers.  The additional moir\'e effects are secondary to the variation of adjacent layer tunneling, but do have a quantitative impact on moir\'e superlattice electronic properties of TTLG systems.

Using this platform, we examine the dependence of moir\'e electronic structure on the accidental lateral stacking shift $\bd_M$, which is not controlled experimentally in current devices.  For the interesting case of middle-layer-twist devices, we view the electronic structure through the filter of mirror symmetry, which plays an essential role in the ideal $\bd_M=0$ (AdA stacked) case. In such a configuration a crossing between a strongly dispersive odd-parity band and even-parity bands emerges. The latter show magic-angle flatband behavior at a twist angle that is larger than the case in twisted bilayer by a factor of $\sqrt{2}$. We characterize electronic structure not only by bandstructure plots and DOS profiles, but also by calculations of the Drude weight, the Hall conductivity, and the Hall density - which are closely related to routine transport characterizations of moir\'e superlattices.  We find that all electronic properties can be strongly altered by accidental lateral stacking shifts. Our results for the Hall density and the Drude weight can be used to experimentally infer the value of $\bd_M$ in a particular device.

Recent experimental work\cite{Park_2021} has demonstrated that AdA trilayers at their magic angle can exhibit superconducting domes similar to those of magic-angle twisted bilayer graphene (MATBG).  The simplest interpretation of these experiments is that superconductivity occurs in the even parity bands, with the odd-parity bands acting as passive spectators, and that the microscopic physics behind the superconductivity is very much like that of MATBG.  If so, our work shows that $\bd_M$ must be small in the measured devices.  This property could be confirmed by Hall density measurements, for which our calculations show that odd-parity bands are not spectators at $\bd_M=0$.  It is possible that the value of $\bd_M$ achieved by the stacking processes employed today is not accidental, and that $\bd_M = 0 $ is preferred for unknown reasons.  In this case superconductivity in magic angle twisted bilayers and trilayers could indeed be very similar.  Our analysis motivates Hall density and Drude weight measurments that could 
prove or disprove this Ansatz.

Moir\'e superlattice continuum models in multilayers are less constrained by symmetry than in the bilayer case.  Our mircoscopic DFT calculations capture details that have not been included in previous continuum models. Although the secondary terms in the model are less important than tunneling between adjacent layers, they do have a significant effect on observable quantities. Their importance is magnified by the narrow widths of the flatbands, especially for the symmetry protected cases of $\bd_M=0$ middle-layer twisted trilayers. For this case, the energetic alignment of flat bands and the decoupled Dirac cone strongly affects the DOS and the Hall conductivities.  As our quantitative understanding of graphene multilayer moir\'e systems improves, 
including these secondary terms will play a more important role.

Many of these conclusions concerning middle-layer twisted trilayers generalize to odd-number-layer twisted multilayers, which are projected to feature flat bands at even larger twist angles, and therefore possibly a higher superconducting $T_c$.  Understanding accidental lateral stacking shift properties and the relationship between lateral shift variability and stacking processes will be important if we wish to use layer number as an intentional design parameter 
for multilayer moir\'es.

\section{Acknowledgements}
Work at University of Texas at Austin was supported by DOE grant DE- FG02-02ER45958. L.L. and F.L. acknowledge support from the Austrian Science Fund (FWF), project I-3827 and the Austrian Marshall Plan Foundation. The authors acknowledge the Texas Advanced Computing Center (TACC) at The University of Texas at Austin for providing HPC resources that have contributed to the research results reported in this paper.

\appendix

\setcounter{table}{0}
\renewcommand{\thetable}{A\arabic{table}}

\section{Definitions for pristine graphene}
\label{app:pristine}
Throughout the manuscript the following lattice parameters are assumed: The lattice vectors of a pristine unit cell of graphene are defined as,
\begin{equation}\label{latvec}
  \textbf{a}_1 = a(0,-1) ~~ \textbf{a}_2 = a(\frac{\sqrt{3}}{2},\frac{1}{2})
\end{equation}
with $a = 2.46$ the lattice constant (except for DFT calculations). With the corresponding reciprocal lattice vectors as,
\begin{equation}\label{replatvec}
  \textbf{b}_1 = \frac{2\pi}{a}(\frac{1}{\sqrt{3}},-1) ~~ \textbf{b}_2 = \frac{2\pi}{a}(\frac{2}{\sqrt{3}},0)
\end{equation}
The positions of atoms in the unit cell are: $\bm{\tau}_{_A} = (0,0,0)$ and $\bm{\tau}_{_B} = (a/\sqrt{3},0,0)$.

The Dirac Hamiltonian for a layer rotated by an angle $\theta$ with respect to a fixed coordinate system is
\begin{equation}
  h(\mathbf{k},\theta) = 
    \hbar v_{_F} k\left(
    \begin{array}{cc}
       0 & e^{i(\theta_{\mathbf{k}} -\theta)} \\
        e^{-i(\theta_{\mathbf{k}} -\theta)} & 0 \\
    \end{array} \right).
\end{equation}
The 6 linear combinations of reciprocal lattice vector with smallest magnitude $\{\bG_1,\hdots,\bG_6\}$ are defined as,
\begin{eqnarray}
  \mathbf{G}_1 &=& \frac{4\pi}{\sqrt{3}a}(\frac{1}{2},-\frac{\sqrt{3}}{2}), ~~
  \mathbf{G}_2 = \frac{4\pi}{\sqrt{3}a}(-\frac{1}{2},\frac{\sqrt{3}}{2}), ~~\\
  \mathbf{G}_3 &=& \frac{4\pi}{\sqrt{3}a}(-1,0), ~~
  \hspace{0.45cm}\mathbf{G}_4 = \frac{4\pi}{\sqrt{3a}}(-\frac{1}{2},-\frac{\sqrt{3}}{2}), ~~\\
  \mathbf{G}_5 &=& \frac{4\pi}{\sqrt{3}a}(\frac{1}{2},\frac{\sqrt{3}}{2}), ~~
  \hspace{0.25cm}\mathbf{G}_6 = \frac{4\pi}{\sqrt{3}a}(1,0). ~~
\end{eqnarray}
Fourier components of coupling elements $\hat{t}^{ll'}_{\alpha\beta,\mathbf{G}}$ between aligned layers and equally Fourier components of the spacial variation of site energies $\hat{\epsilon}^{l}_{\alpha,G}$, which allow the construction of 
\begin{equation}
t^{ll'}_{\alpha\beta}(\br) = \sum_{\bG=\{ 0,\bG_1,\hdots,\bG_6\} } \hat{t}^{ll'}_{\alpha\beta,\mathbf{G}} e^{-i\bG \cdot \bd(\br)}
\end{equation}
and 
\begin{equation}
    \epsilon^{l}_{\alpha}(\br) = \sum_{\bG=\{0,\bG_1,\hdots,\bG_6\}} \hat{\epsilon}^{l}_{\alpha,\bG} e^{-i\bG \cdot \bd(\br)}
\end{equation}
are listed in Table~\ref{tbl:intralayer_coupling_elements}-\ref{tbl:aligned_coupling_elements}. The tables list the zeroth ($\hat{f}_{0}$) and first ($\hat{f}_{\bG_{1,\hdots,6}}$) intralayer (see Table~\ref{tbl:intralayer_coupling_elements}) and interlayer (see Table~\ref{tbl:aligned_coupling_elements}) Fourier component. All other components can be deduced from phase relations listed in the table captions. 

\begin{table}
    \begin{tabular}{c|c|c|c|c|c}
          $\hat{t}^{bb}_{AB,0}$ & 
          $\hat{t}^{bb}_{AB,\bG_1}$&
          $\hat{t}^{mm}_{AB,0}$ & 
          $\hat{t}^{mm}_{AB,\bG_1}$&
          $\hat{t}^{tt}_{AB,0}$ & 
          $\hat{t}^{tt}_{AB,\bG_1}$ \\
          $\hat{\epsilon}_{A,0}^{b}$ &
          $\hat{\epsilon}_{A,\bG_1}^{b}$&
          $\hat{\epsilon}_{A,0}^{m}$ & 
          $\hat{\epsilon}_{A,\bG_1}^{m}$&
          $\hat{\epsilon}_{A,0}^{t}$ & 
          $\hat{\epsilon}_{A,\bG_1}^{t}$ \\ 
          \hline \hline
        \multicolumn{6}{c}{AAd} \\ \hline
          2611 & 0 & 2606 & 2(60) & 2605 & 2(60) \\
          6 & 0 & -9 & 2(28) & 3 & 2(-31) \\
          \hline
        \multicolumn{6}{c}{ABd} \\ \hline
          2591 & 0 & 2608 & 2(60) & 2606 & 2(60) \\
          2 & 0 & -13 & 2(-40) & 11 & 2(30) \\
          \hline
        \multicolumn{6}{c}{AdA} \\ \hline
          2605 & 2(60) & 2602 & 3(60) & 2605 & 2(60) \\
          5 & 2(-30) & -10 & 3(30) & 5 & 2(-30) \\
          \hline
        \multicolumn{6}{c}{AdB} \\ \hline
          2605 & 2(60) & 2602 & 1 & 2583 & 2(-60) \\
          5 & 1(-60) & -11 & 2(-36) & 6 & 1(-150) \\
          \hline
    \end{tabular}\\
    \caption{Fourier components of intralayer coupling constants. Units are set to meV, subscripts b,m,t label the bottom, middle and top layer respectively, phases are denoted in brackets. In this table the Fourier components not explicitly listed $\hat{f}_{\bG_{2,\hdots,6}}$ can be obtained via $\hat{f}_{\bG_1} = \hat{f}_{\bG_3} = \hat{f}_{\bG_5} = \hat{f}^{\dagger}_{\bG_2} = \hat{f}^{\dagger}_{\bG_4} = \hat{f}^{\dagger}_{\bG_6}$ if $\alpha=\beta$ and $\hat{f}_{\bG_1} = \hat{f}_{\bG_4} = \hat{f}_{\bG_2} e^{i 2\pi/3} = \hat{f}_{\bG_5} e^{i2\pi/3} = \hat{f}_{\bG_4}  e^{-i2\pi/3} = \hat{f}_{\bG_6}  e^{-i2\pi/3}$ if $\alpha \neq \beta$ \cite{Jeil2014}. Furthermore $\hat{\epsilon}^{l}_{A}=\hat{\epsilon}^{l,\dagger}_{B}$ and $\hat{t}_{AB}=\hat{t}_{BA}^{\dagger}$.}
    \label{tbl:intralayer_coupling_elements}
\end{table}
\begin{table}
     \begin{tabular}{c|c|c|c}
          $\hat{t}^{bm}_{AA,0}$ & 
          $\hat{t}^{bm}_{AA,\bG_1}$ & 
          $\hat{t}^{bm}_{BB,0}$ & 
          $\hat{t}^{bm}_{BB,\bG_1}$ \\ 
          $\hat{t}^{bm}_{AB,0}$ & 
          $\hat{t}^{bm}_{AB,\bG_1}$ &
          $\hat{t}^{bm}_{BA,0}$ & 
          $\hat{t}^{bm}_{BA,\bG_1}$ \\ 
          \hline \hline
        \multicolumn{4}{c}{AAd} \\ \hline
          225 & 0 & 225 & 0 \\
          0 & 0 & 0 & 0 \\
          \hline
        \multicolumn{4}{c}{ABd} \\ \hline
          0 & 0 & 0 & 0 \\
          357 & 2 & 0 & 0 \\
          \hline
    \end{tabular} \hspace{0.3cm}
    \begin{tabular}{c|c|c|c} 
          $\hat{t}^{bt}_{AA,0}$ & 
          $\hat{t}^{bt}_{AA,\bG_1}$ &
          $\hat{t}^{bt}_{BB,0}$ & 
          $\hat{t}^{bt}_{BB,\bG_1}$ \\
          $\hat{t}^{bt}_{AB,0}$ & 
          $\hat{t}^{bt}_{AB,\bG_1}$ &
          $\hat{t}^{bt}_{BA,0}$ & 
          $\hat{t}^{bt}_{BA,\bG_1}$ \\ 
          \hline \hline
        \multicolumn{4}{c}{AdA} \\ \hline
          4 & 2(-70) & 4 & 2(70)\\
          0 & 2(60) & 0 & 2(-60)\\
          \hline
        \multicolumn{4}{c}{AdB} \\ \hline
          0 & -2 & 0 & 2(-60)\\
          4 & 2(-60) & 0 & -2\\
          \hline
    \end{tabular}
    \caption{Fourier components of coupling constants between aligned layers. Units are set to meV, subscripts b, m, t label the bottom, middle and top layer respectively, phases are denoted in brackets. In this table the Fourier components not explicitly listed $\hat{f}_{\bG_{2,\hdots,6}}$ can be obtained via $\hat{f}_{\bG_1} = \hat{f}_{\bG_3} e^{i2\pi/3} = \hat{f}_{\bG_5} e^{-i2\pi/3} = \hat{f}^{\dagger}_{\bG_4} = \hat{f}^{\dagger}_{\bG_2} e^{i2\pi/3} = \hat{f}^{\dagger}_{\bG_6} e^{-i2\pi/3}$ if $\alpha=\beta$ and $\hat{f}_{\bG_1} = \hat{f}_{\bG_3} e^{-i2\pi/3} = \hat{f}_{\bG_5} e^{i2\pi/3} = \hat{f}^{\dagger}_{\bG_4} = \hat{f}^{\dagger}_{\bG_2} e^{-i2\pi/3} = \hat{f}^{\dagger}_{\bG_6} e^{2i\pi/3}$ if $\alpha=A$ and $\beta=B$. If $\alpha=B$ and $\beta=A$, the relations $\hat{f}_{\bG_1} = \hat{f}_{\bG_3} = \hat{f}_{\bG_5}  = \hat{f}_{\bG_2}^{\dagger} = \hat{f}_{\bG_4}^{\dagger} = \hat{f}_{\bG_6}^{\dagger}$ hold.} 
    \label{tbl:aligned_coupling_elements}
\end{table}

\section{DFT Calculations}\label{ab_intio_details}
All DFT calculations were performed using the Vienna ab initio simulation package(VASP)~\cite{vasp}, within the  local density approximation and the associated equilibrium lattice constant of $a = 2.449\rm{\AA}$ (note that this deviates from the values used elsewhere in the manuscript) and a periodic image separation of 25$\rm{\AA}$. We used a Monkhorst k-point grid of $25\times25\times1$. We sampled the configuration space on a $10\times10\times1$ grid via successive lateral translations of the corresponding twisted layer by $\frac{a}{10}$ along each lattice vector direction. We performed 100 pristine trilayer graphene calculations for each stacking configuration. At each stacking configuration we allowed atomic positions to equilibrate along the out-of-plane axis. Subsequently we transformed and truncated (to $p_z$-orbitals) each resulting pristine cell Hamiltonian into real space using Wannier90 \cite{Marzari1997_wannier,Souza2001_wannier}. To obtain explicit parameters for the continuum model, we deduce effective Fermi velocities and Fermi alignments using calculations where interlayer coupling elements are neglected.

\section{Drude Weight}\label{drude}

In the absence of disorder, the Drude weight tensor of a metal \cite{Kohn1964},
\begin{equation}
  D_{\mu\nu} = 
    \pi \lim_{\omega \rightarrow 0} 
    \omega \rm{Im} \sigma_{\mu\nu}(\omega),
\end{equation}
is related \cite{Resta2018} to the ac dependent conductivity
($\sigma_{\mu\nu}^{reg}(\omega)$) by 
\begin{equation}
    \sigma_{\mu\nu}(\omega) = 
    D_{\mu\nu}[\delta(\omega) +
    \frac{i}{\pi\omega}] + 
    \sigma_{\mu\nu}^{reg}(\omega).
\end{equation}
In the case of band metal, with Kohn’s formula\cite{Kohn1964}, the Drude weight becomes the Fermi volume integral:
\begin{equation}
    D_{\mu\nu} = 2\pi e^2 
    \sum_{n}\int_{BZ}\frac{d\bk}{(2\pi)^d}
    \frac{\partial f_{\bk}}
    {\partial \varepsilon} 
    \frac{1}{\hbar^2}
    \frac{\partial^2\varepsilon_{n\bk}}
    {\partial k_{\mu} \partial k_{\nu}}, 
\end{equation}
with $ \partial f_{\bk}/\partial \varepsilon \equiv \partial f(\varepsilon_{n\bk},\mu)/\partial \varepsilon_{n\bk} $ where $ f(\varepsilon_{n\bk},\mu) $ is the Fermi-Dirac occupation function, $\mu$ is the chemical potential, $\hbar$ is the reduced Planck constant and $ \varepsilon_{n\bk} $ is the band energy. By means of an integration by parts, the Drude weight in two-dimensions may be calculated using band velocities, via:
\begin{equation}\label{drude_formula_apendix}
  D_{\mu\nu} = \frac{e^2}{2\pi} 
    \sum_{n}\int_{BZ}d\bk
    \frac{\partial f_{\bk}}
    {\partial \varepsilon} 
    v_{n\mu} v_{n\nu},
\end{equation}
where the velocity $v_{n\mu}$ may be calculated with the velocity operator $v_{n\mu} = \langle n\bk |\partial H/ \partial k_{\mu} |n\bk \rangle $. In the case of $C_{3v}$ symmetry, the tensor of Drude weight must include this symmetry according to Neumann's principle \cite{Malgrange2014}. The representation of $C_{3v}$ symmetry is:
\begin{equation}
    R(2\pi/3) = \begin{pmatrix}
                 \cos \frac{2\pi}{3} & -\sin \frac{2\pi}{3} & 0 \\
                 \sin \frac{2\pi}{3} & \cos \frac{2\pi}{3} & 0 \\
                 0 & 0 & 1 \\
                \end{pmatrix}.
\end{equation}
With the tensor of Drude weight after the operation of $R(2\pi/3)$ denoted as $D'_{\mu\nu}$, the invariance of the components of the tensor implies that $D'_{\mu\nu} = D_{\mu\nu}$, that is:
\begin{equation}
    D'_{\mu\nu} = a_{\mu \alpha} a_{\nu \beta} D_{\alpha \beta} = D_{\mu\nu},
\end{equation}
using Einstein’s convention, with $a_{\mu \alpha}$ and $a_{\nu \beta}$ the matrix element of $R(2\pi/3)$. The tensor of Drude weight with $C_{3v}$ symmetry is thus:
\begin{equation}
    D = \begin{pmatrix}
                 D_{xx} & 0 & 0 \\
                 0 & D_{xx} & 0 \\
                 0 & 0 & D_{zz} \\
                \end{pmatrix}.
\end{equation}

\bibliography{ttlg}
\end{document}